\documentclass[10pt, a4]{article}
\usepackage[a4paper,bottom=1in,top=1in,left=1in,right=1in]{geometry}
\usepackage{graphicx}
\usepackage{amsmath}
\usepackage{amssymb}

\usepackage{hyperref}
\hypersetup{
     colorlinks 		= true,
     linkcolor 		= blue,
     anchorcolor 		= blue,
     citecolor 		= blue,
     filecolor 		= blue,
     urlcolor 		= blue
     }

\newcommand*{\figref}[2][]{%
  \hyperref[{fig:#2}]{%
    \ref*{fig:#2}%
    \ifx\\#1\\%
    \else
      \,#1%
    \fi
  }%
}

\newcommand{\citet}[1]{\cite{#1}}

\newcommand{\beq}{\begin{equation}}
\newcommand{\eeq}{\end{equation}}
\newcommand{\eps}{\varepsilon}
\newcommand{\yt}{\tilde{y}}
\newcommand{\ut}{\tilde{u}}
\newcommand{\vt}{\tilde{v}}
\newcommand{\wt}{\tilde{w}}
\newcommand{\pt}{\tilde{p}}

\newcommand{\olY}{\overline{Y}}

\newcommand{\olv}{\overline{V}}

\newcommand{\ola}{\overline{\alpha}}
\newcommand{\olb}{\overline{\beta}}
\newcommand{\olg}{\overline{\gamma}}

\begin{document}

\title{Stability of the flow due to a linear stretching sheet}

\author{P.~T.~Griffiths$^{1,}$\thanks{Email: \href{mailto:paul.griffiths@coventry.ac.uk}{paul.griffiths@coventry.ac.uk}}\,\,, S.~O.~Stephen$^{2}$ and M.~Khan$^{3}$
\\
\\
$^{1}$Centre for Fluid and Complex Systems, Coventry University,
\\
Coventry CV1 5FB, United Kingdom
\\
\\
$^{2}$School of Mathematics and Statistics, The University of Sydney,
\\
Sydney NSW 2006, Australia
\\
\\
$^{3}$Department of Mathematics, Quaid-i-Azam University,
\\
Islamabad 45320, Pakistan}

\date{}

\noindent This article has been accepted for publication in Physics of Fluids. The DOI number for this manuscript is \href{https://aip.scitation.org/doi/pdf/10.1063/5.0060645}{doi: 10.1063/5.0060645}

\vspace{\baselineskip}

\noindent Cite as: Phys.~Fluids \textbf{33}, 084106 (2021)

\vspace{\baselineskip}

\begin{abstract}
In this article we consider the linear stability of the two-dimensional flow induced by the linear stretching of a surface in the streamwise direction. The basic flow is a rare example of an exact analytical solution of the Navier-Stokes equations. Using results from a large Reynolds number asymptotic study and a highly accurate spectral numerical method we show that this flow is linearly unstable to disturbances in the form of Tollmien-Schlichting waves. Previous studies have shown this flow is linearly stable. However, our results show that this is only true for G\"{o}rtler-type disturbances.
\end{abstract}

\maketitle

\section{\label{sec:intro}Introduction}

Boundary-layer flows induced by the extrusion of a surface have received considerable attention since they were first described by \citet{Sakiadis1961}. Interest in these types of flows stems not only from the fact that they often admit analytical solutions to the Navier-Stokes equations; they are also used to model a variety of industrial processes. These so called `stretching' flows are of practical importance to chemical and metallurgy industries where extrusion processes are commonplace. One such example would be the manufacture of a polymer sheet being continuously extruded from a casting die. Bounded stretching and/or shrinking surface flows also have a number of physiological applications and can be used to model transmyocardial laser revascularisation (TMLR), see, for example \citet{Waters2001}.

These types of boundary-layer flows exhibit fluid entrainment and are therefore qualitatively different to Blasius-type boundary layer flows. In fact, two- and three-dimensional stretching flows are much more closely aligned to stagnation-point flows. The specific focus of this article will be on flows induced by an impermeable sheet with a velocity that increases linearly in the direction of stretching. \citet{Vleggaar1977} has shown that this linear behaviour is appropriate in relevant experiments. This specific case of linear stretching was first considered by \citet{Crane1970} who derived an exact analytical solution of the Navier-Stokes equations in two dimensions. The equivalent three-dimensional problem was first considered by \citet{Wang1984} who demonstrated that although exact analytical solutions could not be obtained, the problem could be solved numerically by introducing a suitable similarity approach. Given that these flows exhibit fluid entrainment at the boundary-layer edge, and that the surface stretching is linear in nature, an association can be made between these flows and the three-dimensional flow induced by a rotating disk. Very recently the ideas of Crane and Wang have been extended by \citet{Ayatsetal2021} to include the family of flows associated with the independent stretching or shrinking of two infinite parallel plates.  

A wealth of literature exists concerning flows over different types of stretching surfaces. Crane's analysis of the flow induced by a linear sheet velocity has been extended to consider power-law type sheet velocities by \citet{Vleggaar1977}, and exponential type sheet velocities by \citet{MagyariKeller1999}. \citet{GuptaGupta1977}, included the effects of both permeability and heat and mass transfer demonstrating that exact analytical solutions for the temperature variation across the layer can be obtained in the case when the Prandtl number is equal to unity. In the cases when this condition does not hold the solutions can be generalised and are described in terms of the incomplete gamma function. The relatively recent article by \citet{Al-HousseinyStone2012} provides a chronological summary of the literature on boundary-layer flows due to stretching impermeable sheets and the interested reader is referred to this manuscript for a more comprehensive list of previous studies. In addition to the studies noted by \citet{Al-HousseinyStone2012}, there are many more studies that consider these types of flows with the addition of magnetohydrodynamic (\citet{ChakrabartiGupta1979}), non-Newtonian (\citet{Rajagopaletal1984}) and pressure gradient (\citet{RileyWeidman1989}) effects. There are yet more studies that consider combinations of these effects for permeable/impermeable stretching (or indeed shrinking) sheets, one particularly nice example is Liao's investigation \cite{Liao2003}, which details a range of analytical solutions for magnetohydrodynamic non-Newtonian stretching flows.

Given the vast number of studies detailing basic flow solutions for a plethora of different stretching configurations, it is surprising just how little attention the associated stability problems have received. \citet{BhattacharyyaGupta1985}, were the first to investigate the linear stability of the two-dimensional Crane solution, determining that the flow is linearly stable to infinitesimal G\"{o}rtler-type disturbances. Seeking disturbances of this nature is an entirely justifiable procedure since the flow exhibits streamline curvature. It is well understood that G\"{o}rtler-type disturbances form owing to centrifugal instability. This is in contrast to Tollmien-Schlichting (TS) waves that are observed in flows that exhibit zero curvature at the wall, such as the Blasius boundary-layer, and the stretching sheet configuration consider herein. Indeed, \citet{BhattacharyyaGupta1985} note that ``\dots\textit{our stability analysis is confined to infinitesimal
G\"{o}rtler-type disturbances, which are non-propagating. It cannot, therefore, be ruled out that the flow may be unstable to other types of disturbances which may be infinitesimal or of finite amplitude}''. Bhattacharyya and Gupta's analysis was then extended by \citet{Takharetal1989} to include magnetohydrodynamic effects. The authors note that the magnetic field has a stabilising influence on a flow that was already linearly stable. In much the same fashion, \citet{Dandapatetal1994} consider an extension to include viscoelastic fluid effects, determining that, for disturbance wavelengths shorter than the viscoelastic lengthscale, the flow will again be further stabilised.

More recently \citet{DavisPozrikidis2014} have revisited the linear stability of the flow induced by a linear stretching sheet. In a departure from the analysis of \citet{BhattacharyyaGupta1985}, the authors stipulate that the streamwise dependence of the disturbances must match that of the base state. Again, searching for G\"{o}rtler-type disturbances, they find that the flow is linearly stable to both two- and three-dimensional disturbances. A far-field asymptotic analysis details a remarkably simple relationship between the growth rate of the disturbance and the disturbance wavenumber. The work associated in the derivation of this relationship offers an insight as to why this flow is not susceptible to this type of disturbance.

In this study we will analyse the stability of Crane's solution from an alternative viewpoint. Rather than seeking disturbances of G\"{o}rtler-type we will instead investigate the possibility of the development of small-amplitude perturbations in the form of TS waves. TS instability waves are a prominent feature of many flat plate boundary-layer flows and their growth has been shown to be well described by linear stability theory. It is clear from the statements of Bhattacharyya and Gupta that the importance of these types of disturbances in stretching flows has been known about for some time. Within this investigation we will outline both asymptotic and numerical stability analyses that show that Crane's flow is, in fact, linearly \textit{unstable} to small-amplitude TS waves.

In \ref{subsec:form1} we formulate the problem and derive the governing perturbation equations. In \ref{subsec:form2} we describe our numerical scheme in detail and validate it against the results of \citet{DavisPozrikidis2014}. In \ref{subsec:form3} we derive the appropriate energy balance equations for problems of this type. Having completed these derivations and our validation exercise, in \ref{sec:num_results}, we present linear stability results for two- and three-dimensional disturbances and also results from our integral energy analysis, which provides insights as to the mechanisms responsible for this observed instability. In \ref{sec:asym} our numerical results are compared to high Reynolds number asymptotic predictions, where we focus primarily on the most unstable case, 2D perturbations. Excellent agreement is observed between our asymptotic and numerical findings. Lastly, in \ref{sec:disc}, our results are discussed and placed in context and potential avenues for further work are outlined.

\section{\label{sec:form}Problem Formulation and Validation of the Numerical Scheme}

\subsection{\label{subsec:form1}Derivation of the basic state and the governing perturbation equations}

We consider the flow of a steady, incompressible, Newtonian fluid induced by the stretching of an infinite planar surface. The streamwise, wall-normal and spanwise coordinates are $x^{*}$, $y^{*}$, and $z^{*}$, respectively. The associated fluid velocities in these directions are then $\tilde{\textbf{U}}^{*}=(\tilde{U}^{*},\tilde{V}^{*},\tilde{W}^{*})$. The surface stretching acts along the $y^{*}=0$ plane and is linear in nature. Note that throughout this analysis an asterisk indicates a dimensional quantity. 

The system is governed by the following conservation of mass and momentum equations

\begin{subequations}
\begin{align}
\nabla^{*}\cdot\tilde{\textbf{U}}^{*}&=0,
\label{gova}
\\
\frac{\textrm{D}\tilde{\textbf{U}}^{*}}{\textrm{D}t^{*}}&=-\frac{1}{\rho^{*}}\nabla^{*}\tilde{P}^{*}+\nu^{*}\Delta^{*}\tilde{\textbf{U}}^{*},
\label{govb}
\end{align}
\label{gov}
\end{subequations}

\noindent where $t^{*}$ is time, $\rho^{*}$ is the fluid density, $\tilde{P}^{*}$ is pressure and $\nu^{*}$ is the kinematic viscosity of the fluid. The two-dimensional mean flow (see Fig.~\figref[(a)]{flow}) is solved subject to the following boundary conditions

\begin{equation}
\tilde{U}^{*}(y^{*}=0)-\xi^{*}x^{*}=\tilde{V}^{*}(y^{*}=0)=0,\quad\tilde{U}^{*}(y^{*}\to\infty)\to0,
\label{gov_bound}
\end{equation}

\noindent where the constant $\xi^{*}$ is the stretching rate with units $\textrm{s}^{-1}$. The exact solution of \eqref{gov} subject to \eqref{gov_bound} was first reported by \citet{Crane1970}. By introducing the following similarity variables

\begin{equation*}
U(y)=\frac{\tilde{U}^{*}}{\xi^{*}x^{*}},\quad V(y)=\frac{\tilde{V}^{*}}{\sqrt{\xi^{*}\nu^{*}}},\quad P(y)=\frac{\tilde{P}^{*}}{\rho^{*}\xi^{*}x^{*}},
\end{equation*}

\noindent where $y=y^{*}/L^{*}$, and the non-dimensionalising length-scale is $L^{*}=\sqrt{\nu^{*}/\xi^{*}}$, the governing equations are reduced to the following set of coupled non-linear ODEs

\begin{align*}
U+V'&=0,
\\
U^{2}+VU'&=U'',
\\
VV'&=-P'+V''.
\end{align*}

\noindent These must be solved subject to 

\begin{equation*}
U(y=0)-1=V(y=0)=0,\quad U(y\to\infty)\to0.
\end{equation*}

\begin{figure}[t!]
\centering
\includegraphics[width=75mm]{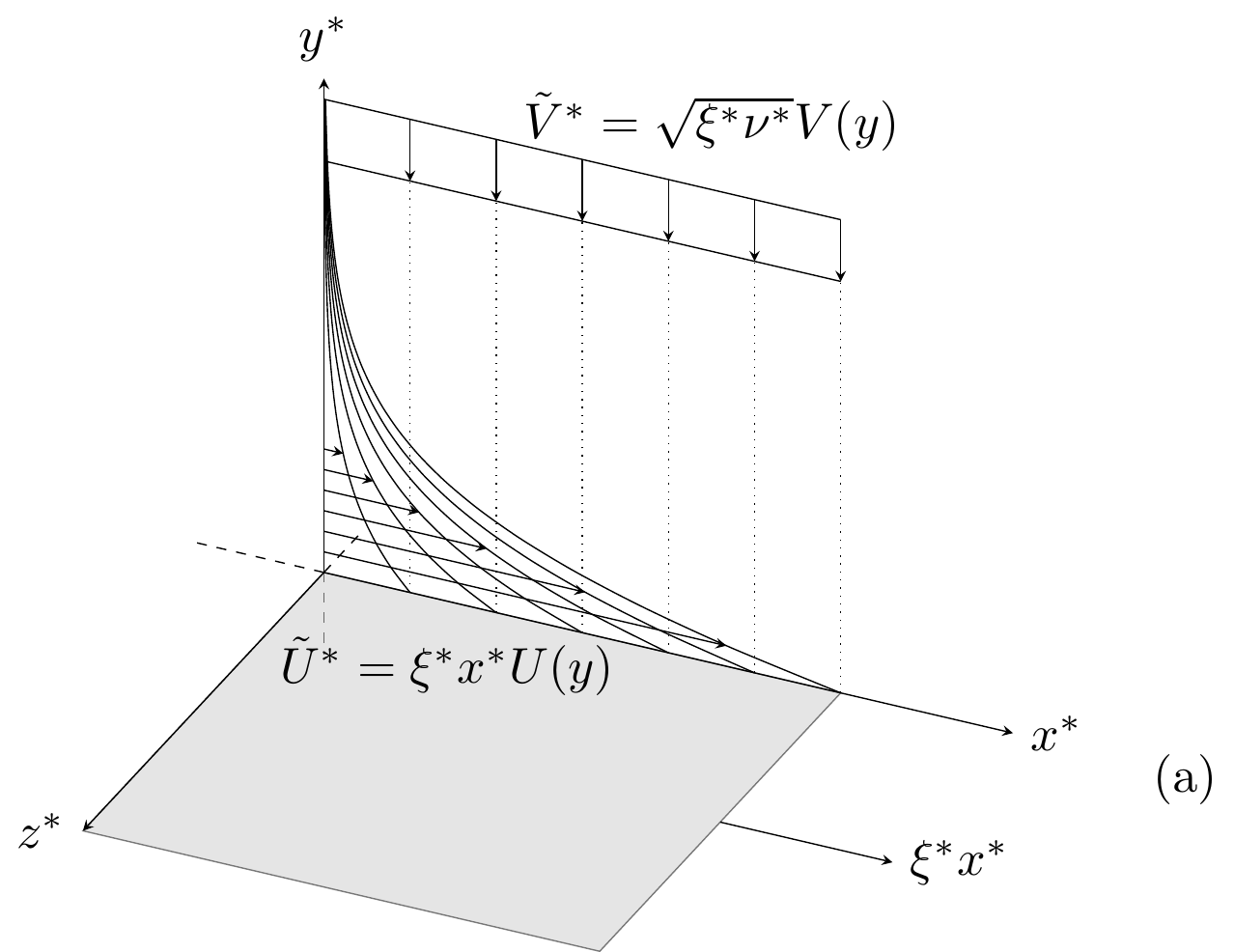}
\includegraphics[width=75mm]{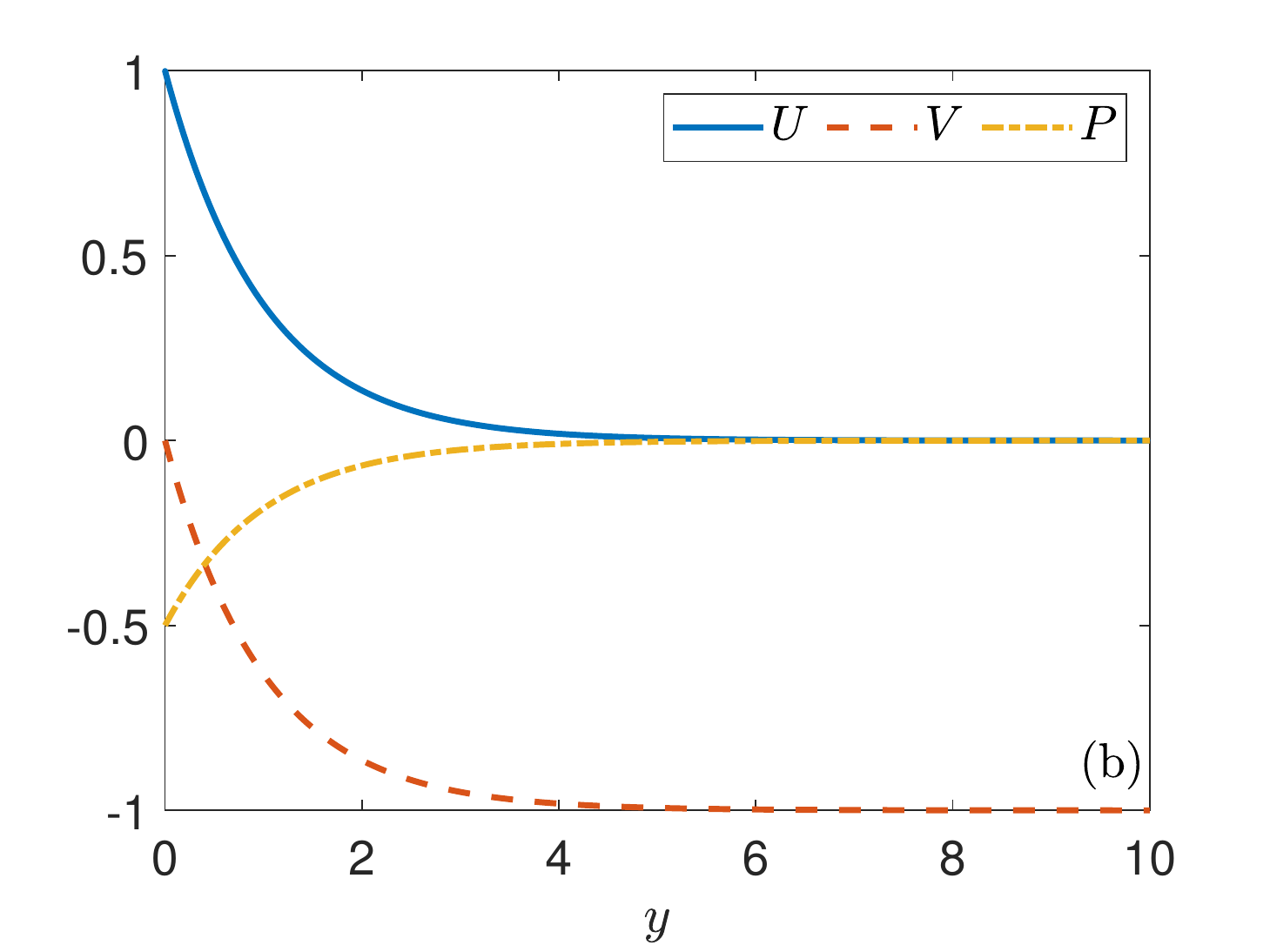}
\caption{\label{fig:flow} Basic flow profiles induced by a linear stretching sheet. A schematic diagram of the flow is presented in $\textrm{(a)}$. The streamwise profile $\tilde{U}^{*}=\xi^{*}x^{*}U(y)$, linearly increases with the streamwise coordinate. The wall-normal flow component is independent of the streamwise coordinate and, far away from the wall, is directed towards the stretching surface with uniform velocity $\tilde{V}_{\infty}^{*}=-\sqrt{\xi^{*}\nu^{*}}$. The exact similarity profiles are presented in $\textrm{(b)}$. To ensure that the reference pressure at the edge of the boundary layer is zero, the constant $P_{0}$ is set equal to minus one-half.}
\end{figure}

\noindent As noted by \citet{Crane1970}, the streamwise and wall-normal velocities and pressure have a remarkably simple analytical solution 

\begin{equation*}
U=\textrm{e}^{-y},\quad V=\textrm{e}^{-y}-1,\quad P=P_{0}+\frac{(1-\textrm{e}^{-2y})}{2},
\end{equation*}

\noindent where $P_{0}=P(y=0)$. These solutions are highlighted in Fig.~\figref[(b)]{flow}. The stability analysis is then applied at a streamwise location $x_{s}^{*}$. The local Reynolds number is $R=x_{s}^{*}\xi^{*}L^{*}/\nu^{*}=x_{s}^{*}/L^{*}=x_{s}$. Thus, the local Reynolds number is identically equivalent to a dimensionless streamwise location along the stretching sheet. In this sense, this flow is representative of the two-dimensional equivalent of the flow due to a rotating disk where one finds that a radial location along the disk is equivalent to the local Reynolds number \cite{Lingwood1995}. The non-dimensionalising velocity, pressure and time-scales are then $x_{s}^{*}\xi^{*}$, $\rho^{*}(x_{s}^{*}\xi^{*})^{2}$ and $L^{*}/(x_{s}^{*}\xi^{*})$. 

\newpage

\noindent After non-dimensionalisation the mean flow quantities are perturbed as follows

\begin{alignat*}{2}
\tilde{U}(x,y,z,t)&=\frac{x}{R}U(y)&&+u(x,y,z,t),
\\
\tilde{V}(x,y,z,t)&=\frac{1}{R}V(y)&&+v(x,y,z,t),
\\
\tilde{W}(x,y,z,t)&=&&+w(x,y,z,t),
\\
\tilde{P}(x,y,z,t)&=\frac{1}{R^{2}}P(y)&&+p(x,y,z,t),
\end{alignat*}

\noindent where the perturbation quantities $(u,v,w,p)$ are assumed to be small. The dimensionless governing equations are then linearised with respect to the perturbation quantities. In order to make the perturbation equations separable in $x$, $y$ and $t$, it is necessary to invoke a parallel-flow-type approximation; that is to say that the variable $x$ is replaced throughout by the Reynolds number $R$. This is consistent with the approach adopted by Lingwood \cite{Lingwood1995}, in her corresponding three-dimensional rotating disk analysis, where the spatial variable $r$ is replaced by $R$. Although the flow is not strictly homogeneous in the streamwise direction it is reasonable to make this parallel-flow-type approximation given that we expect the onset of linear instability to occur sufficiently far enough downstream (i.e., at reasonably large Reynolds numbers) from the leading edge. Employing this approximation leads to the retention of the terms associated with the wall-normal mean velocity profile $V$. This is a departure from Blasius-type flat plate boundary-layer studies but is consistent with Lingwood's approach where she also retains terms associated with the wall-normal velocity profile (in her case this is $W$, not $V$, since the flow over a rotating disk is three-dimensional, not two-dimensional). The resulting system of linear perturbation equations is then 

\begin{subequations}
\begin{align}
\nabla\cdot\boldsymbol{u}&=0,\label{pert_gov_a}
\\
\mathcal{L}\boldsymbol{u}+\biggl(U'v+\frac{Uu}{R}\biggr)\boldsymbol{\hat{x}}+\frac{V'v}{R}\boldsymbol{\hat{y}}&
=-\nabla p+\frac{1}{R}\Delta\boldsymbol{u},\label{pert_gov_b}
\end{align}
\label{pert_gov}
\end{subequations}

\noindent where $\boldsymbol{u}=(u,v,w)$, the operator $\mathcal{L}$ is defined like so 

\begin{equation*}
\mathcal{L}=\frac{\partial}{\partial t}+U\frac{\partial}{\partial x}+\frac{V}{R}\frac{\partial}{\partial y},
\end{equation*}

\noindent and $\boldsymbol{\hat{x}}$ and $\boldsymbol{\hat{y}}$ are the unit vectors in the streamwise and wall-normal directions, respectively. At this stage it is worth noting that the last two terms on the left-hand side of \eqref{pert_gov_b}, and the final term associated with the operator $\mathcal{L}$, appear due to the $x$-dependence of the streamwise velocity component and the non-zero nature of the wall-normal velocity component. The retention of these terms is entirely consistent with the three-dimensional analysis presented by \cite{Lingwood1995} and ensures that the action of surface stretching is considered explicitly. These terms are not present in the standard linearised governing equations, see, for example, the analysis presented by \citet{SchmidHenningson2001}.

If the perturbations are assumed to have the normal mode form

\begin{equation*}
(u,v,w,p)=[\hat{u}(y),\hat{v}(y),\hat{w}(y),\hat{p}(y)]\textrm{e}^{\textrm{i}(\alpha x+\beta z-\omega t)},
\end{equation*}

\noindent then \eqref{pert_gov} reduces to the following system of coupled ODEs

\begin{equation}
\left\{
\frac{\alpha^{2}}{R}
\begin{bmatrix}
1 & 0 & 0 & 0 \\ 
0 & 1 & 0 & 0 \\ 
0 & 0 & 1 & 0 \\
0 & 0 & 0 & 0
\end{bmatrix}
+
\textrm{i}\alpha
\begin{bmatrix}
U & 0 & 0 & 1 \\ 
0 & U & 0 & 0 \\ 
0 & 0 & U & 0 \\
1 & 0 & 0 & 0
\end{bmatrix}
+
\begin{bmatrix}
\mathcal{S}_{+1} & \mathcal{D}U & 0 & 0 \\ 
0 & \mathcal{S}_{-1} & 0 & \mathcal{D} \\ 
0 & 0 & \mathcal{S}_{0} & \textrm{i}\beta \\
0 & \mathcal{D} & \textrm{i}\beta & 0
\end{bmatrix}
\right\}
\begin{bmatrix}
\hat{u} \\ 
\hat{v} \\ 
\hat{w} \\
\hat{p}
\end{bmatrix}
=
\begin{bmatrix}
0 \\
0 \\
0 \\
0
\end{bmatrix},
\label{ODEs}
\end{equation}

\noindent where $\mathcal{S}_{n}=(R^{-1}\beta^{2}-\textrm{i}\omega)+R^{-1}(V\mathcal{D}-\mathcal{D}^{2})+nR^{-1}U$, and $\mathcal{D}=\textrm{d}/\textrm{d}y$. In our spectral analysis we assume that the frequency of the disturbance $\omega$, and the spanwise wavenumber $\beta$, are solely real whilst the streamwise wavenumber $\alpha$, is assumed to be complex $\alpha=\alpha_{\textrm{r}}+\textrm{i}\alpha_{\textrm{i}}$. As such, disturbances will grow exponentially in space when $\alpha_{\textrm{i}}<0$.

Combining the governing equations together gives an Orr-Sommerfeld-type equation for this problem

\begin{equation}
(\mathcal{D}^{2}-\gamma^{2})^{2}\hat{v}-\textrm{i}\alpha R[(U-c)(\mathcal{D}^{2}-\gamma^{2})\hat{v}-\hat{v}\mathcal{D}^{2}U]=V(\mathcal{D}^{2}-\gamma^{2})\mathcal{D}\hat{v}-U(\mathcal{D}-\gamma^{2})\hat{v}+\textrm{i}\beta\mathcal{D}(U\hat{w}),
\label{Orr}
\end{equation}

\noindent where $\gamma^{2}=\alpha^{2}+\beta^{2}$, and $c=\omega/\alpha$. The terms on the right-hand side are due to the influence of the surface stretching and do not normally appear in the standard Orr-Sommerfeld equation. The appearance of the spanwise wavenumber in \eqref{Orr} means that it is not possible to ensure that Squire's theorem will hold in this case. As such, it will be necessary to investigate both 3D and 2D modes in this study.

Given that the perturbation velocities are subject to the no-slip condition at the wall, and that all perturbations must decay to zero far from the surface, \eqref{ODEs} is solved subject to

\begin{align}
\hat{u}(y=0)=\hat{v}(y=0)=\hat{v}'(y=0)=\hat{w}(y=0)=0,&
\nonumber
\\
\hat{u}(y\to\infty)\to\hat{v}(y\to\infty)\to\hat{w}(y\to\infty)\to\hat{p}(y\to\infty)\to0.&
\label{bcs_ODEs}
\end{align}

\noindent The condition on the derivative of the wall-normal perturbation at $y=0$, is simply a consequence of the form of the continuity equation.

\subsection{\label{subsec:form2}Numerical Method and Validation}

The system of linear ODEs \eqref{ODEs} is solved subject to the above boundary conditions \eqref{bcs_ODEs} using a spectral method that utilises Chebyshev polynomials. These polynomials are defined like so

\begin{equation*}
T_{n}(Y_{j})=\cos[n\arccos(Y_{j})],\quad n=0,1,2,\dots,N-1, N,
\end{equation*}

\noindent where $N$ is the number of collocation points and the Gauss-Lobatto points $Y_{j}$, are defined as such

\begin{equation*}
Y_{j}=-\cos\biggl(\frac{j\pi}{N}\biggr),\quad j=0,1,2,\dots,N-1, N.
\end{equation*}

\noindent In order to map the physical domain $0\leq y_{j}\leq y_{\textrm{max}}$, to the computational domain  $-1\leq Y_{j}\leq +1$, the following exponential map is used

\begin{equation*}
y_{j}=-\frac{1}{\phi}\ln\biggl(\frac{1+Y_{j}+A}{A}\biggr),
\end{equation*}

\noindent where $A=2(\textrm{e}^{-\phi y_{\textrm{max}}}-1)^{-1}$, and $\phi$ is a free constant. In order to capture the exponential nature of the base flow a range of different $\phi$ values were tested (see Fig.~\figref[(a)]{num_test}). Our analysis revealed that choosing too large a value for $\phi$ resulted in solutions that did not converge within a desired tolerance, typically $\mathcal{O}(10^{-5})$. This was true irrespective of the choice of $y_{\textrm{max}}$, as the number of collocation points were increased. Too small a value for $\phi$ and the map reverted to a linear approximation, and was therefore unable to capture the exponential decay of the base flow solutions. A more moderate value of the constant, $\phi=1/5$, proved to be a sensible choice, providing fully converged solutions when $y_{\textrm{max}}=40$, and $N=100$. It was somewhat surprising to note the requirement of such a large value for $y_{\textrm{max}}$, in order to obtain accurate numerical solutions. Many corresponding flat plate boundary-layer studies find that converged eigensolutions are obtained when the value of $y_{\textrm{max}}$ is of the order of $20$, for example \citet{Milleretal2018}. However, we note that \citet{DavisPozrikidis2014} arrived at a similar conclusion when conducting their G\"{o}rtler disturbance analysis.  

\noindent In order to validate our numerical scheme it was tested against the results of \citet{DavisPozrikidis2014}. The authors showed, rather remarkably, that in the limit as $y\to\infty$, a simple relationship exists between the growth rate $\lambda^{*}$, and transverse wavenumber $k^{*}$, when searching for G\"{o}rtler-type disturbances

\begin{equation}
\frac{\lambda^{*}}{\xi^{*}}=-(L^{*}k^{*})^{2}-\frac{1}{4}.
\label{Davis}
\end{equation}

\noindent Our spectral scheme, which requires solving a quadratic eigenvalue problem, was used to solve their systems of ODEs subject to the standard conditions of no-slip and no-penetration at the wall and decay in to the far-field. We were able to exactly reproduce the results quoted by the authors, showing excellent agreement with both their asymptotic prediction and the results from their linear eigenvalue temporal analysis, see Fig.~\figref[(b)]{num_test}.

\begin{figure}[t!]
\centering
\includegraphics[width=75mm]{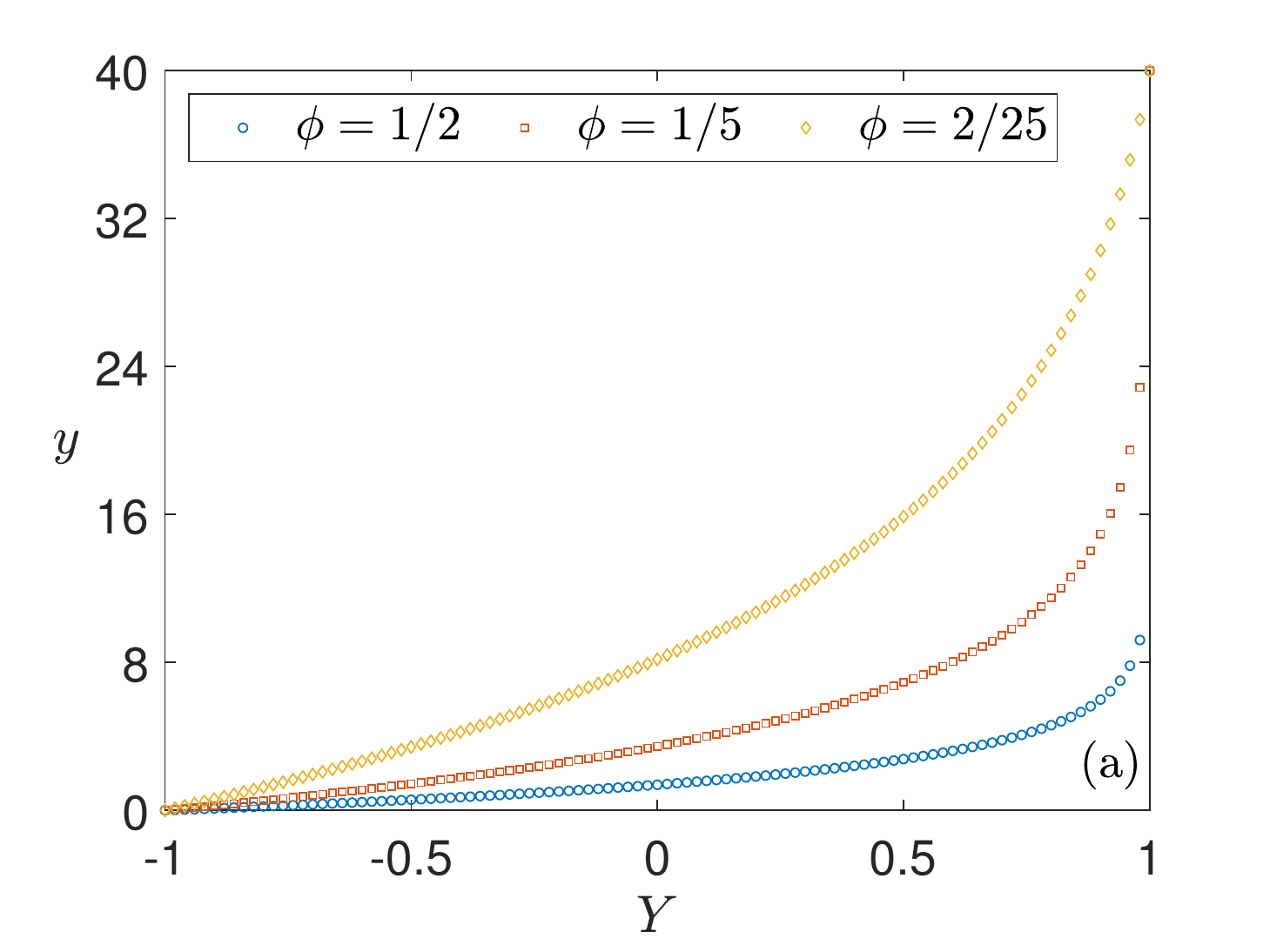}
\includegraphics[width=75mm]{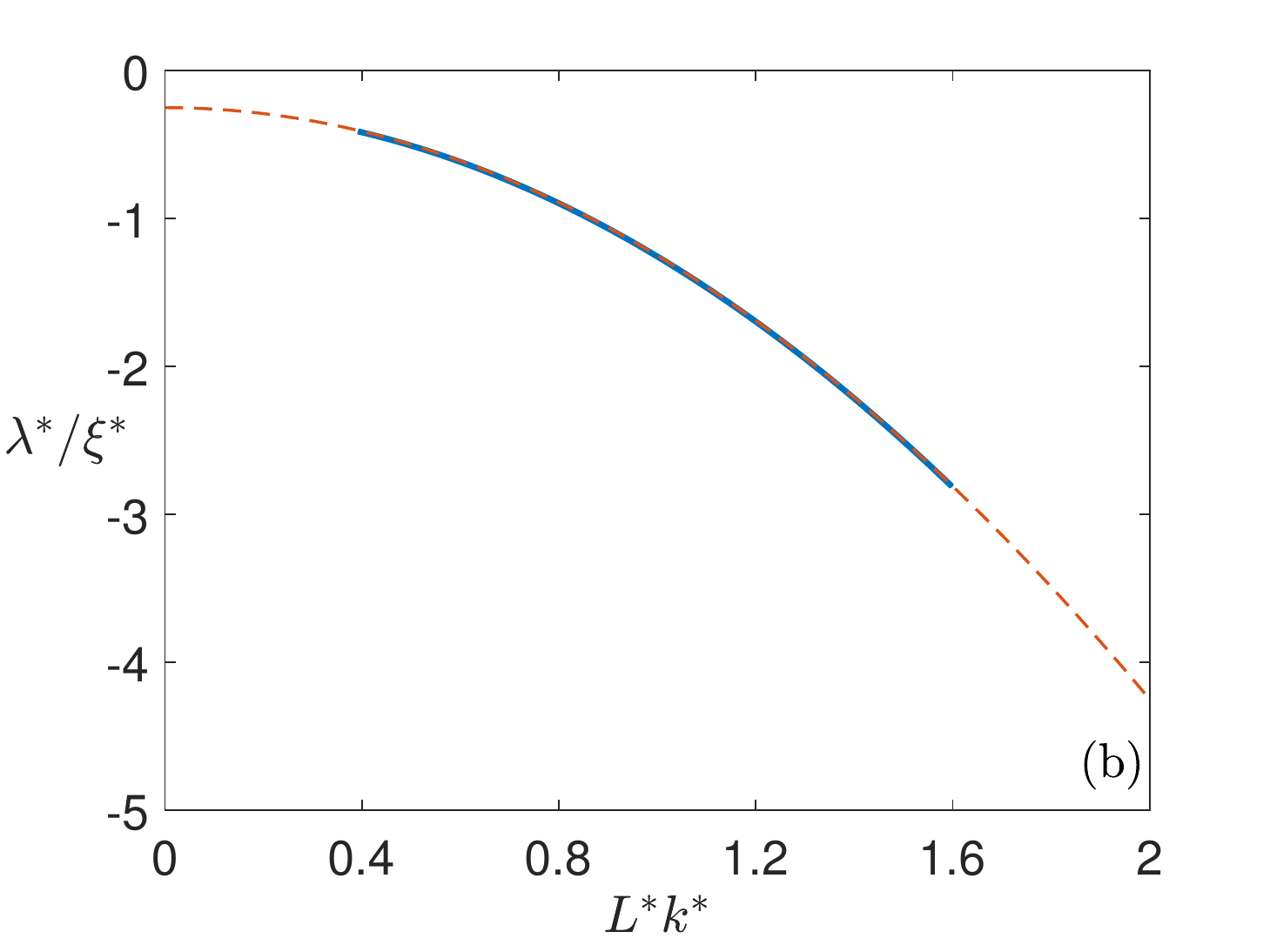}
\caption{\label{fig:num_test} In (a) the physical domain is mapped to the computational domain for a range of different $\phi$ values. In this study we use the value $\phi=1/5$ for all quoted results. In (b) we reproduce the results of \citet{DavisPozrikidis2014}. Our numerical results, in the range $0.4\leq L^{*}k^{*}\leq1.6$, shown in blue, are plotted against the authors asymptotic prediction given in \eqref{Davis}. The growth rate $\lambda^{*}$, is never greater than zero for any value of the transverse wavenumber $k^{*}$, and hence the flow is linearly stable to G\"{o}rtler-type disturbances.}
\end{figure}

\subsection{\label{subsec:form3}Derivation of energy balance equations}

In order to derive an appropriate integral energy analysis for this problem we take a linear combination of the constituent parts of \eqref{pert_gov_b}, average over one time period and integrate across the boundary layer. Having done so we arrive at the governing integral energy equation for flows of this nature

\begin{multline}
\int_{0}^{\infty}\biggl[U\frac{\partial\mathcal{E}}{\partial x}+\frac{\partial(up)}{\partial x}-\frac{1}{R}\frac{\partial(v\Omega_{z}-w\Omega_{y})}{\partial x}\biggr]\,\textrm{d}y+\int_{0}^{\infty}\biggl[\frac{\partial(wp)}{\partial z}+\frac{1}{R}\frac{\partial(v\Omega_{x}-u\Omega_{y})}{\partial z}\biggr]\,\textrm{d}y
\\
=-\int_{0}^{\infty}\biggl[\frac{U(2u^{2}-\mathcal{E})}{R}+uv\mathcal{D}U\biggr]\,\textrm{d}y-\int_{0}^{\infty}\frac{(\Omega_{x}^{2}+\Omega_{y}^{2}+\Omega_{z}^{2})}{R}\,\textrm{d}y,
\nonumber
\end{multline}

\noindent where $\mathcal{E}=(u^{2}+v^{2}+w^{2})/2$, and $\Omega_{i}$ is the $i$-component of vorticity. Given that the perturbations have the normal mode form then it is possible to show that 

\begin{multline}
-2\alpha_{\textrm{i}}\int_{0}^{\infty}\biggl(U\langle\hat{\mathcal{E}}\rangle+\langle\hat{u}\hat{p}\rangle+\overbrace{\frac{\beta\langle\hat{u}\hat{w}\rangle}{R}}^{\textrm{I}}\biggr)\,\textrm{d}y=-\int_{0}^{\infty}\biggl[\frac{U(2\langle\hat{u}^{2}\rangle-\langle\hat{\mathcal{E}}\rangle)}{R}+\langle\hat{u}\hat{v}\rangle\mathcal{D}U\biggr]\,\textrm{d}y
\\
-\int_{0}^{\infty}\biggl[\frac{\langle\hat{u}'^{2}\rangle+\langle\hat{w}'^{2}\rangle}{R}+\underbrace{\frac{(\alpha_{\textrm{r}}^{2}-\alpha_{\textrm{i}}^{2})(\langle\hat{v}^{2}\rangle+\langle\hat{w}^{2}\rangle)}{R}}_{\textrm{II}}+\underbrace{\frac{\beta^{2}(\langle\hat{u}^{2}\rangle+\langle\hat{v}^{2}\rangle)}{R}}_{\textrm{III}}\biggr]\,\textrm{d}y,
\label{full_energy}
\end{multline}

\begin{figure}[t!]
\centering
\includegraphics[width=85mm]{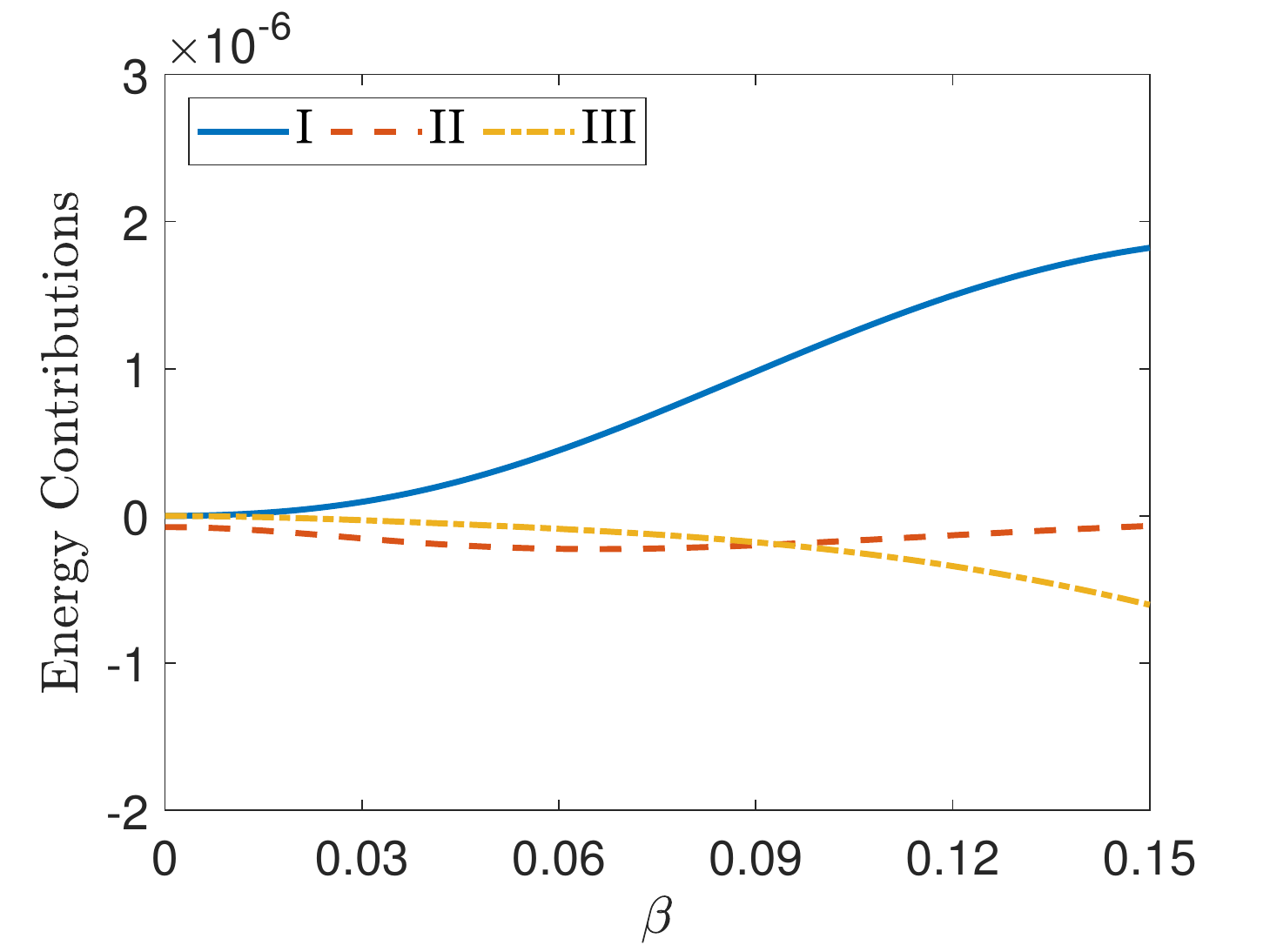}
\caption{\label{fig:energy_test} Variation of the three energy contribution terms highlighted in \eqref{full_energy} with the spanwise wavenumber $\beta$. For each value of $\beta$, the most unstable eigenmode ($\alpha_{\textrm{i}}=\textrm{max}(\alpha_{\textrm{i}})$) is selected, and the integrals as defined in \eqref{full_energy}, are computed at a fixed Reynolds number of $R=1\times10^{5}$. These terms have not been normalised but prove to be negligible and so are excluded from the integral energy analysis and the associated integral energy equation presented in \eqref{energy}.}
\end{figure}

\noindent where $\langle\hat{\mathcal{E}}\rangle=(\langle\hat{u}^{2}\rangle+\langle\hat{v}^{2}\rangle+\langle\hat{w}^{2}\rangle)/2$, and $\langle\hat{x}\hat{y}\rangle=\hat{x}^{\star}\hat{y}+\hat{x}\hat{y}^{\star}$, with $^{\star}$ indicating the complex conjugate. The first term on the right-hand side of \eqref{full_energy} exists solely due to the influence of the surface stretching. The second is the `standard' Reynolds stress term that appears in all 2D and 3D analyses of this type (see, for example, \citet{Porter1998}). All the terms included in the final integral term are associated with the action of viscous dissipation.

Our numerical investigations reveal that for all the cases considered here, terms $\textrm{I}-\textrm{III}$ (highlighted above) are in fact negligible, with their absolute value always being less than $2\times10^{-6}$ (see Fig.~\ref{fig:energy_test}). It transpires that the above integral energy equation can be approximated like so

\begin{equation}
    \underbrace{\vphantom{\int_{0}^{\infty}}
    -2\alpha_{\textrm{i}}}_{\textrm{TME}}\simeq
    \underbrace{-\int_{0}^{\infty}\langle\hat{u}\hat{v}\rangle\mathcal{D}U\,\textrm{d}y}_{\textrm{EPRS}}
	\underbrace{-\int_{0}^{\infty}\frac{U(2\langle\hat{u}^{2}\rangle-
    \langle\hat{\mathcal{E}}\rangle)}{R}\,\textrm{d}y}_{\textrm{EDSS}}
    \underbrace{-\int_{0}^{\infty}\frac{\langle\hat{u}'^{2}\rangle+\langle\hat{w}'^{2}\rangle}{R}\,\textrm{d}y}_{\textrm{EDV}},
    \label{energy}
\end{equation}

\noindent where the right-hand side of \eqref{energy} has been normalised by the integral of the combination of energy flux and the work done by the pressure, $\mathcal{N}=\int_{0}^{\infty}(U\langle\hat{\mathcal{E}}\rangle+\langle\hat{u}\hat{p}\rangle)\,\textrm{d}y$. The energy production term, labelled $\textrm{EPRS}-$ \textbf{E}nergy \textbf{P}roduction due to \textbf{R}eynolds \textbf{S}tresses, will always be positive and the dissipation terms, labelled $\textrm{EDSS}-$ \textbf{E}nergy \textbf{D}issipation due to \textbf{S}urface \textbf{S}tretching and $\textrm{EDV}-$ \textbf{E}nergy \textbf{D}issipation due to \textbf{V}iscosity, will always be negative. Therefore, in the cases when the production is greater than the absolute value of the dissipation, the right-hand side of \eqref{energy} will be greater than zero. In these cases the eigenmode in question is amplified and the \textbf{T}otal \textbf{M}echanical \textbf{E}nergy of the system is positive. Clearly, this can only hold true if $\alpha_{\textrm{i}}<0$, which is consistent with our definition of linear instability.

\section{\label{sec:num_results}Numerical Results}

\begin{table}[t!]
\centering
\caption{\label{tab:critical}Critical values (indicated by the superscript `\textrm{crit}') for the frequency, streamwise wavenumber and Reynolds number for a range of fixed values of the spanwise wavenumber, $\beta$. The critical values of $\omega$ and $\alpha_{\textrm{r}}$ are given to four decimal places whilst the critical Reynolds number is quoted accurate to five significant figures.}
\begin{tabular}{cccc}
\hline\hline
$\beta$	&	$\omega^{\textrm{crit}}$ 	& 	$\alpha_{\textrm{r}}^{\textrm{crit}}$ 	& 	$R^{\textrm{crit}}$	\\
\hline
0       	& 	0.1364 				& 	0.1614 							& 	48499.1			\\
0.05    	& 	0.1310 				& 	0.1550 							& 	50911.5			\\
0.1     	& 	0.1139 				& 	0.1348 							& 	60652.0			\\
0.15    	& 	0.0823 				& 	0.0971	 						& 	96607.5			\\
\hline\hline
\end{tabular}
\end{table}

\begin{figure}[t!]
\centering
\includegraphics[width=75mm]{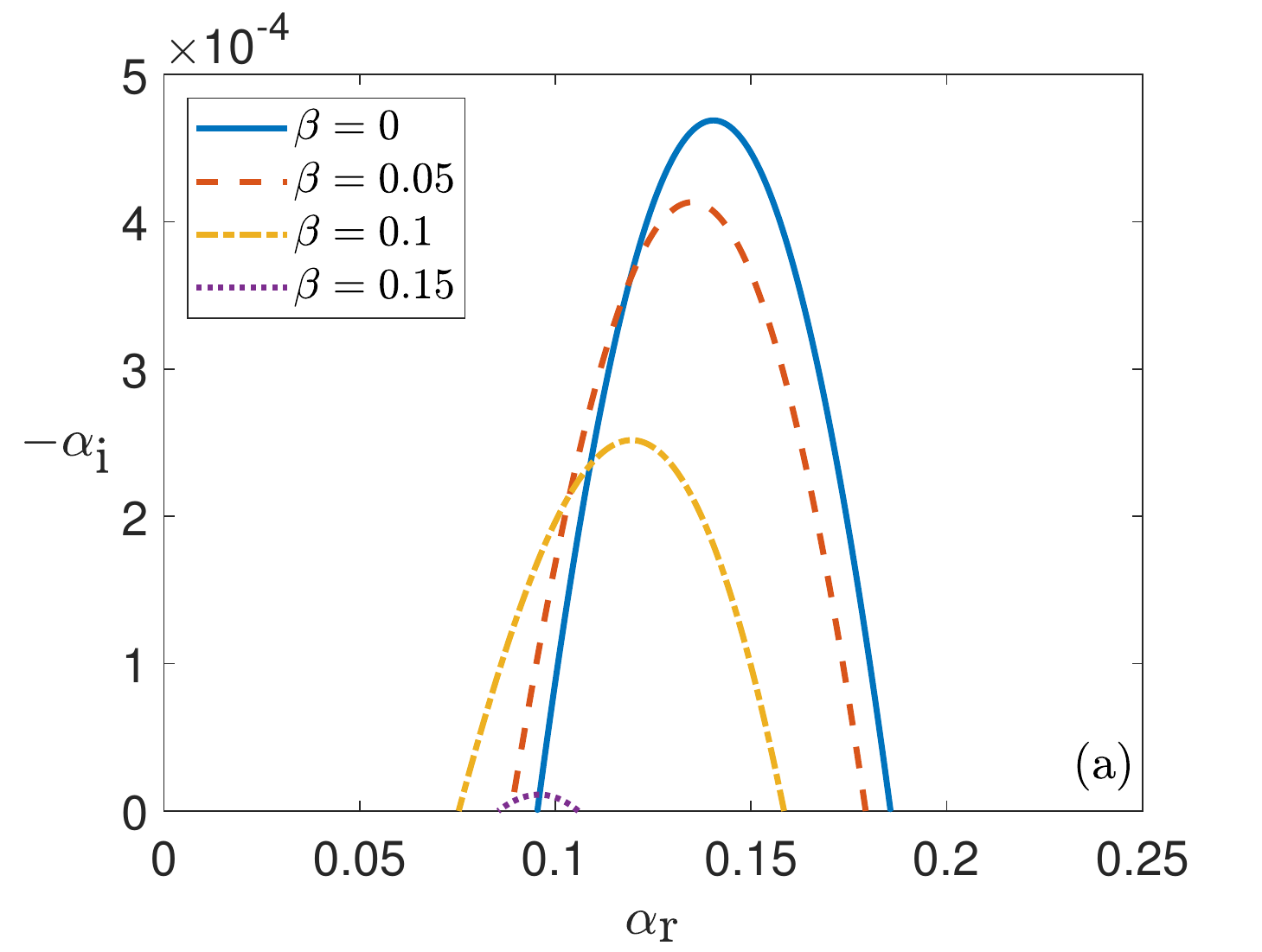}
\includegraphics[width=75mm]{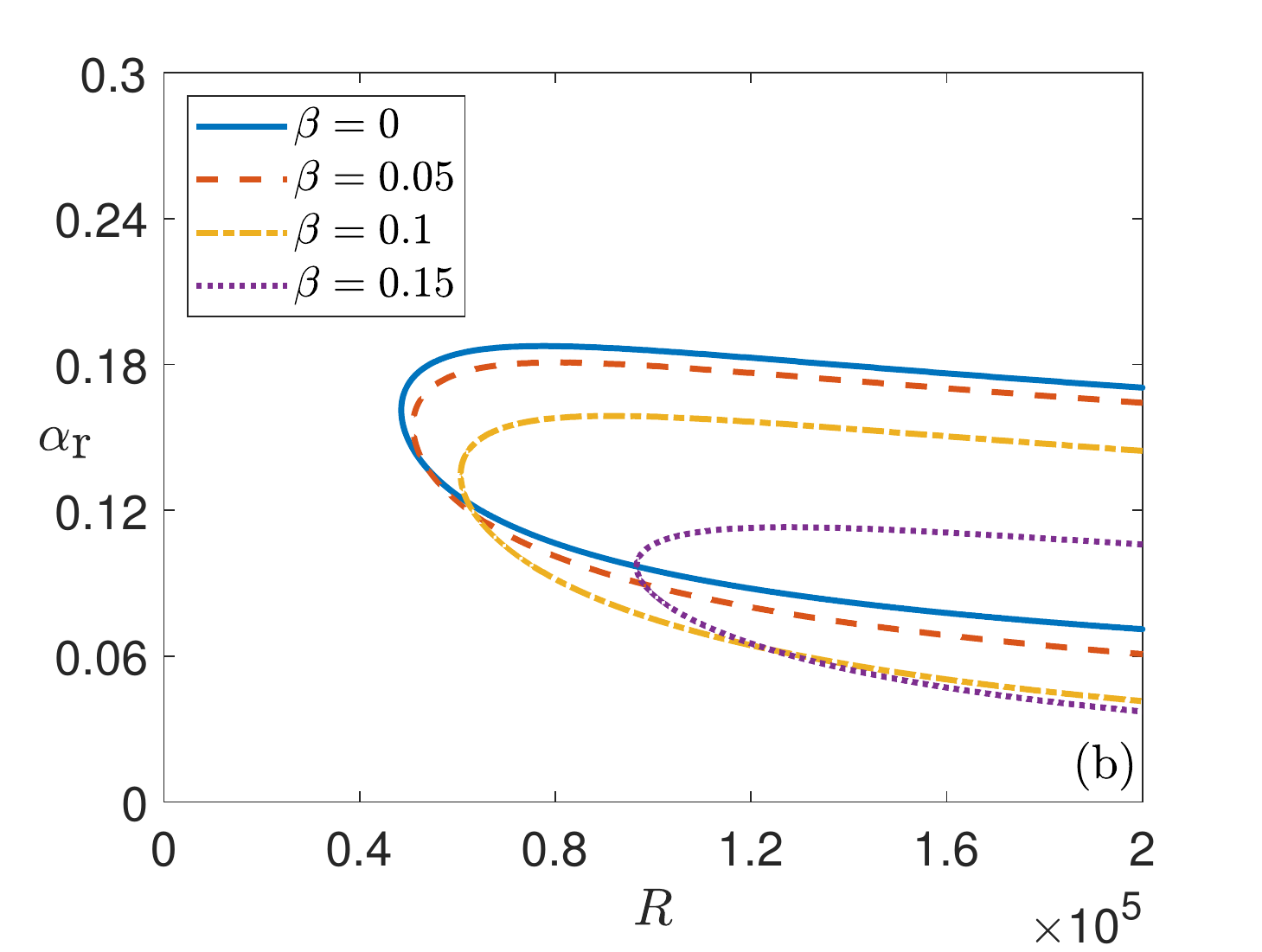}
\caption{\label{fig:num_results} In (a) the growth rate, defined as $-\alpha_{\textrm{i}}$, is plotted against $\alpha_{\textrm{r}}$, for a range of $\beta$ values at a fixed value of the Reynolds number, $R=1\times10^{5}$. In (b) the curves of neutral stability, all the points where $\alpha_{\textrm{i}}=0$, are plotted for the same range of $\beta$ values. As the value of the spanwise wavenumber increases the area encompassed by the curve is reduced.}
\end{figure}

Having validated our numerical scheme we begin by solving \eqref{ODEs} subject to \eqref{bcs_ODEs} for fixed values of $\beta$ whilst cycling through a range of values of $\omega$ and $R$ in order to determine points where $\alpha_{\textrm{i}}\leq0$. A point is deemed to be neutrally stable if $\alpha_{\textrm{i}}=0$. When considering 2D perturbations only ($\beta=0$) we find that the flow is linearly unstable above a critical Reynolds number of $R^{\textrm{crit}}=48,499$ (see Table~\ref{tab:critical}). Although this critical Reynolds number is large when compared to a Blasius-type boundary-layer flow, these results do clearly show that this flow is susceptible to instabilities arising from travelling-wave disturbances. In fact, the critical Reynolds number noted above, is of the same order of that exhibited by other boundary-layer flows with exponentially decaying base flow solutions, for example, the asymptotic suction boundary-layer flow which has a critical Reynolds number, as quoted by \citet{DempseyWalton2017}, of $R^{\textrm{crit}}\simeq54,370$.

Although an equivalent to Squire's theorem cannot be proved for this type of flow we do find that two-dimensional disturbances are indeed the most unstable. Fig.~\figref[(a)]{num_results} presents a snapshot of the linear growth rates for a range of $\beta$ values at a fixed Reynolds number, $R=1\times10^{5}$. We find that the amplitude of the growth rate is significantly reduced as $\beta$ increases. This observed stabilisation is further exemplified by the neutral stability curves for $\alpha_{\textrm{r}}$ presented in Fig.~\figref[(b)]{num_results}, showing that the critical Reynolds number increases as the spanwise wavenumber increases. In addition to this, the area encompassed by the neutral curves is also reduced. Physically, this means that there are fewer wavenumbers that are susceptible to linear instability.

\begin{figure}[t!]
\centering
\includegraphics[width=65mm]{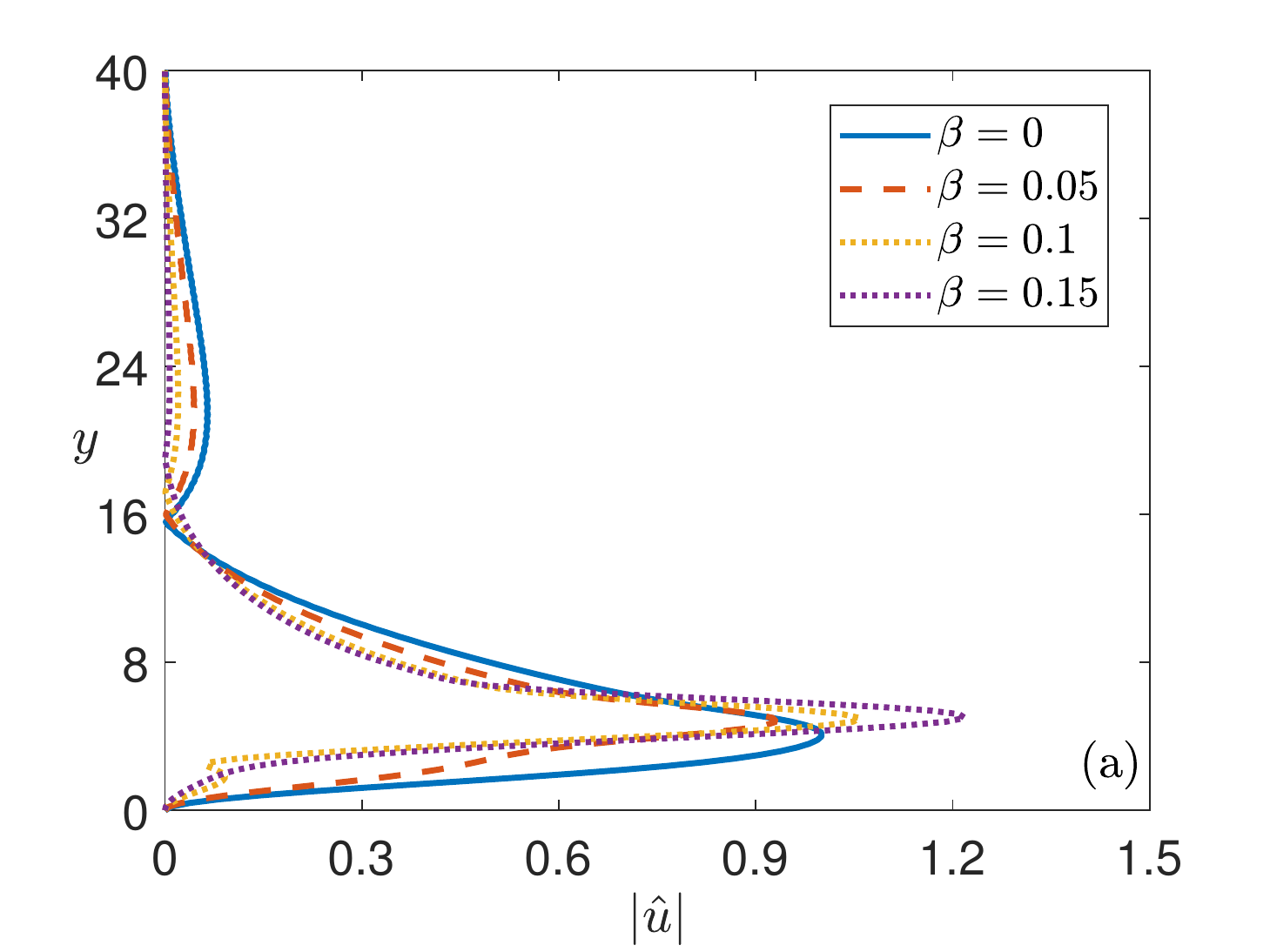}
\includegraphics[width=65mm]{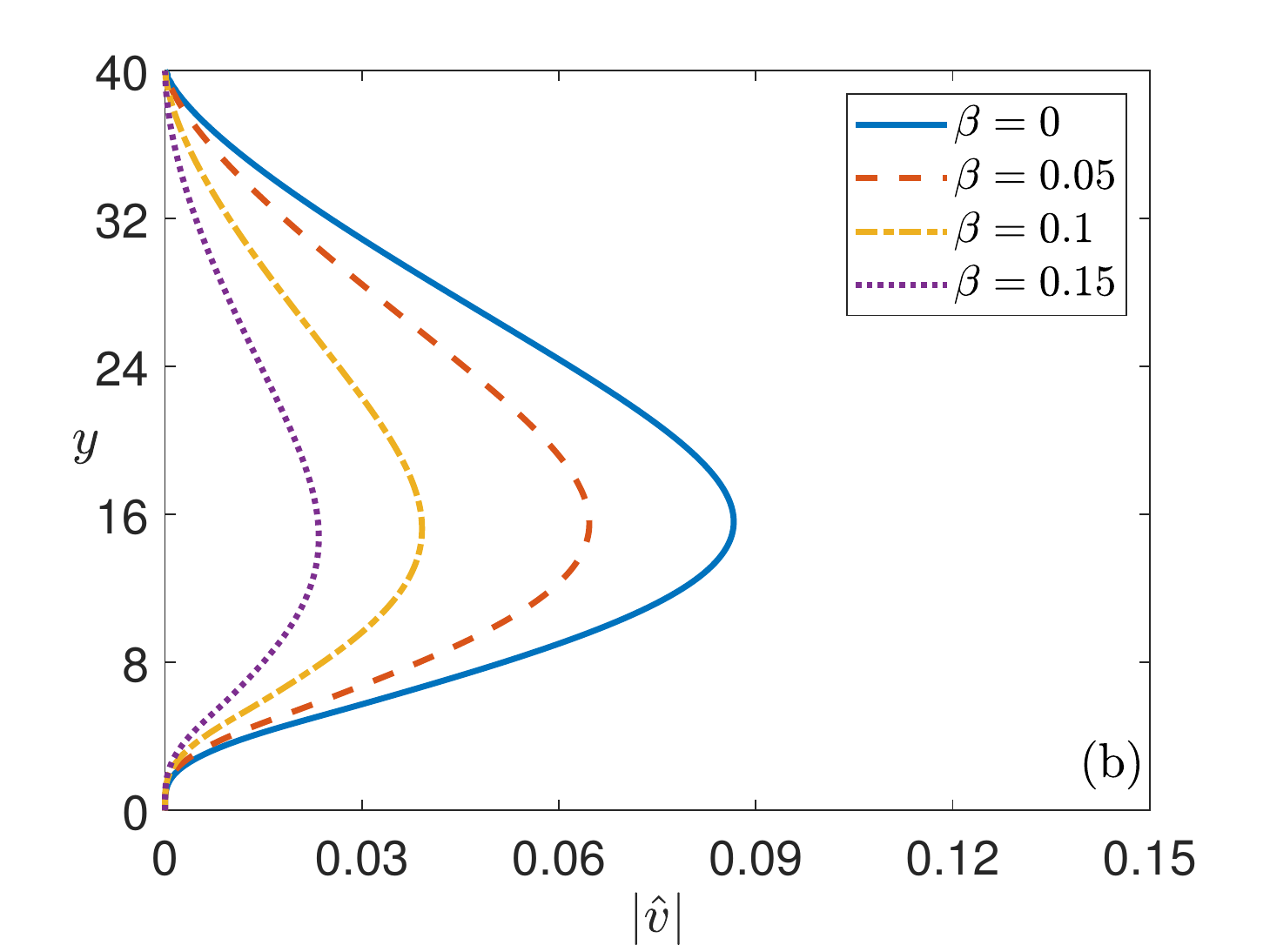}
\includegraphics[width=65mm]{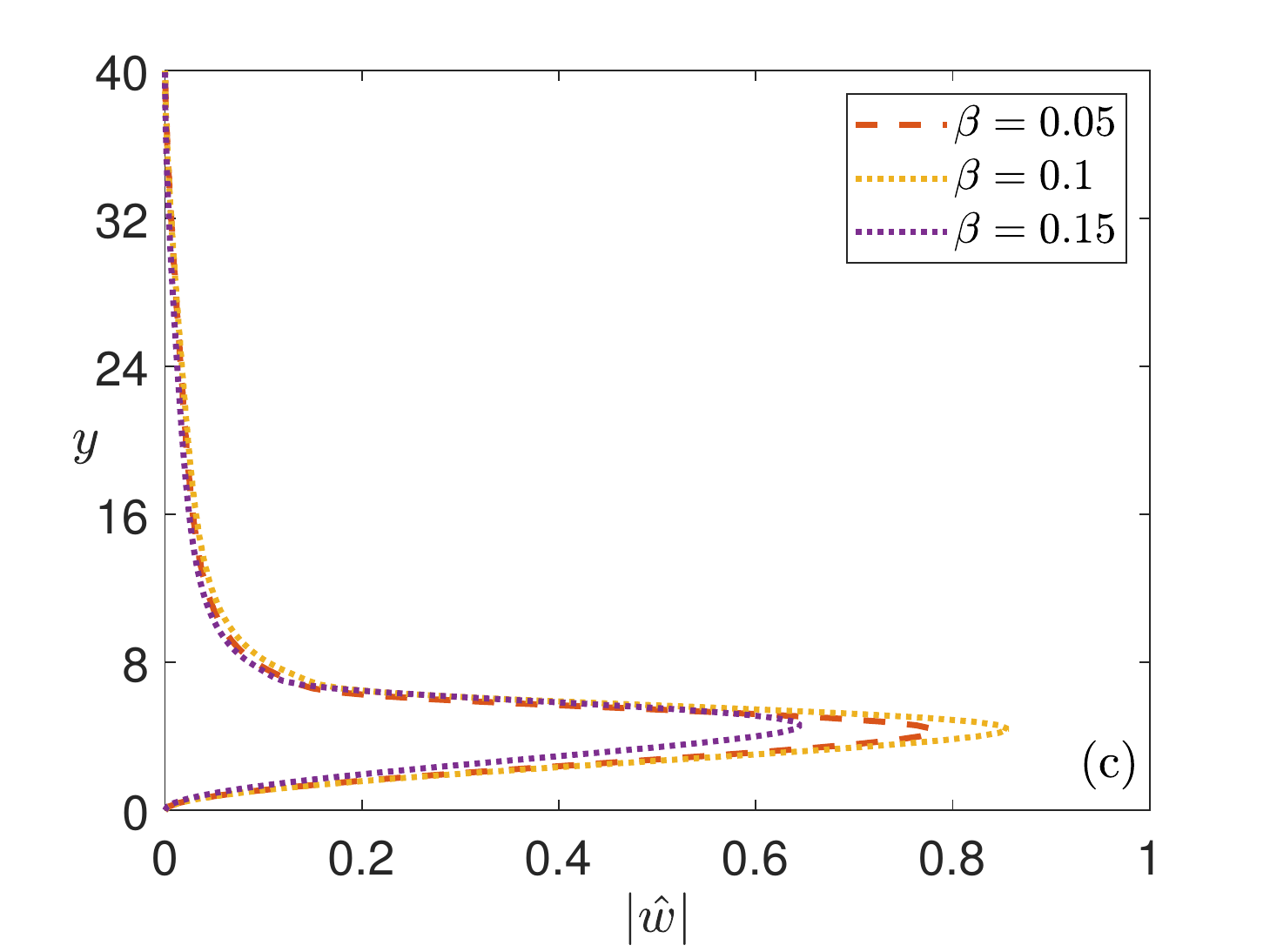}
\caption{\label{fig:eigs} Plots of the streamwise, wall-normal and spanwise eigenfuncions for a range of $\beta$ values at a fixed value of the Reynolds number, $R=1\times10^{5}$. In each case the most unstable eigenmode ($\alpha_{\textrm{i}}=\textrm{max}(\alpha_{\textrm{i}})$) is selected. All the results have been normalised with respect to the maximum value of $|\hat{u}|$ for the case when $\beta=0$.}
\end{figure}

We find that as the value of the spanwise wavenumber is increased above even moderate values, $\beta\gtrapprox0.187$, the flow becomes linearly stable. This suggests that in the cases when $\beta$ is greater than this value the area encompassed by the neutral stability curve becomes vanishingly small, and thus, the critical Reynolds number asymptotes towards positive infinity. Given that we were unable to categorically prove that 3D disturbances would be more stable than 2D disturbances, due to the additional terms appearing in the modified Orr-Sommerfeld equation \eqref{Orr}, these numerical results provide strong evidence that this must indeed be the case.

\begin{figure}[t!]
\centering
\includegraphics[width=65mm]{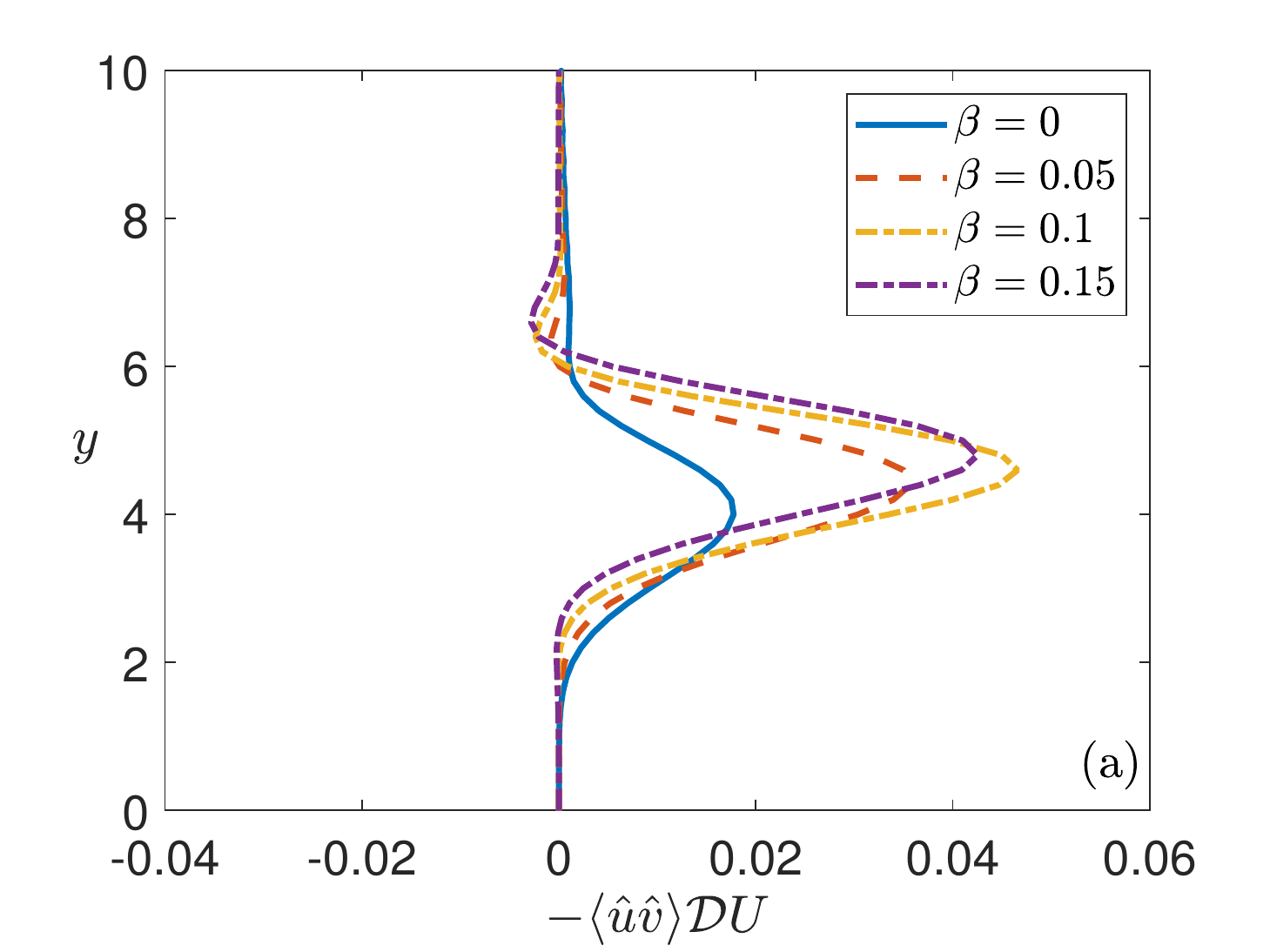}
\includegraphics[width=65mm]{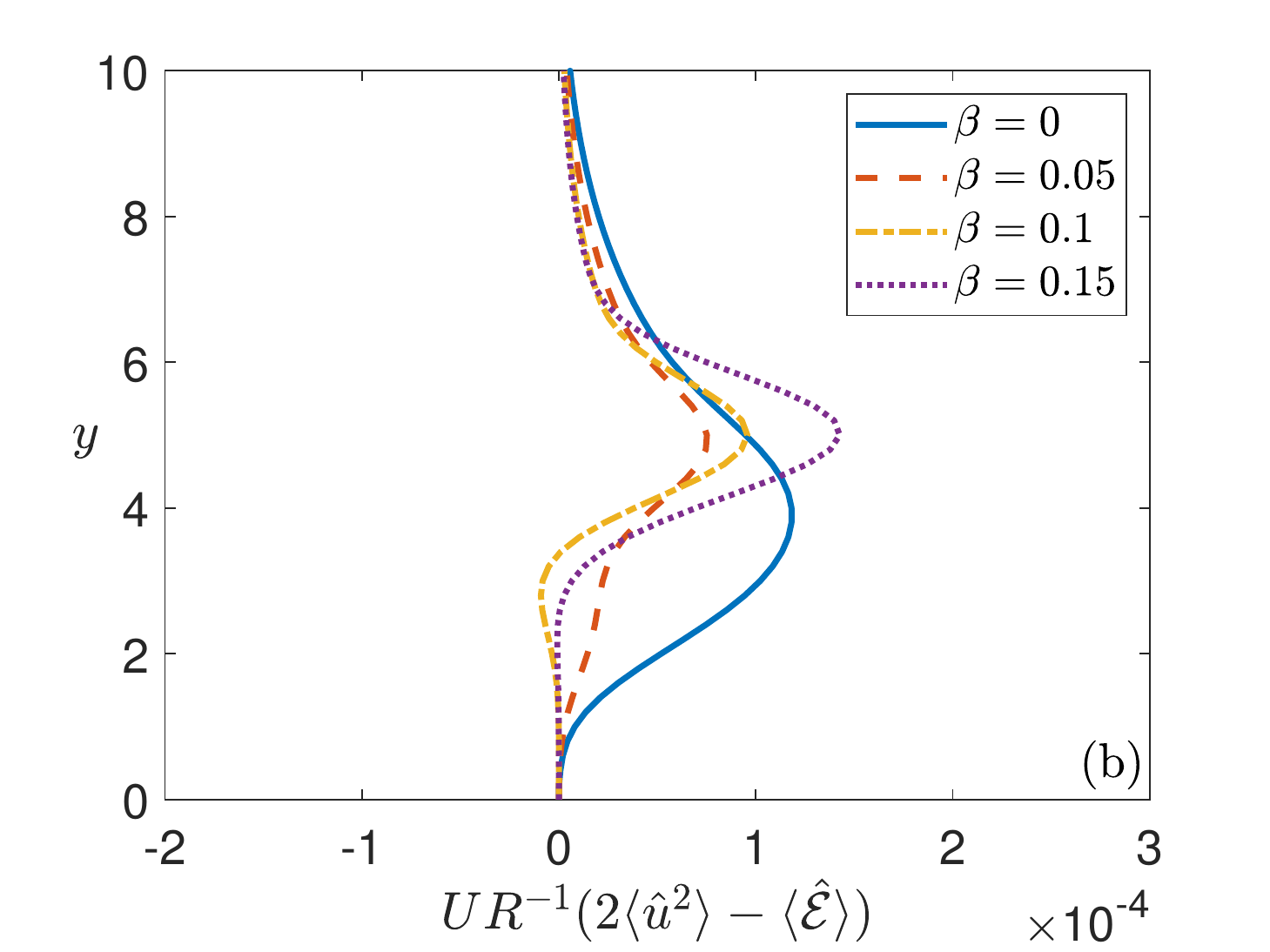}
\includegraphics[width=65mm]{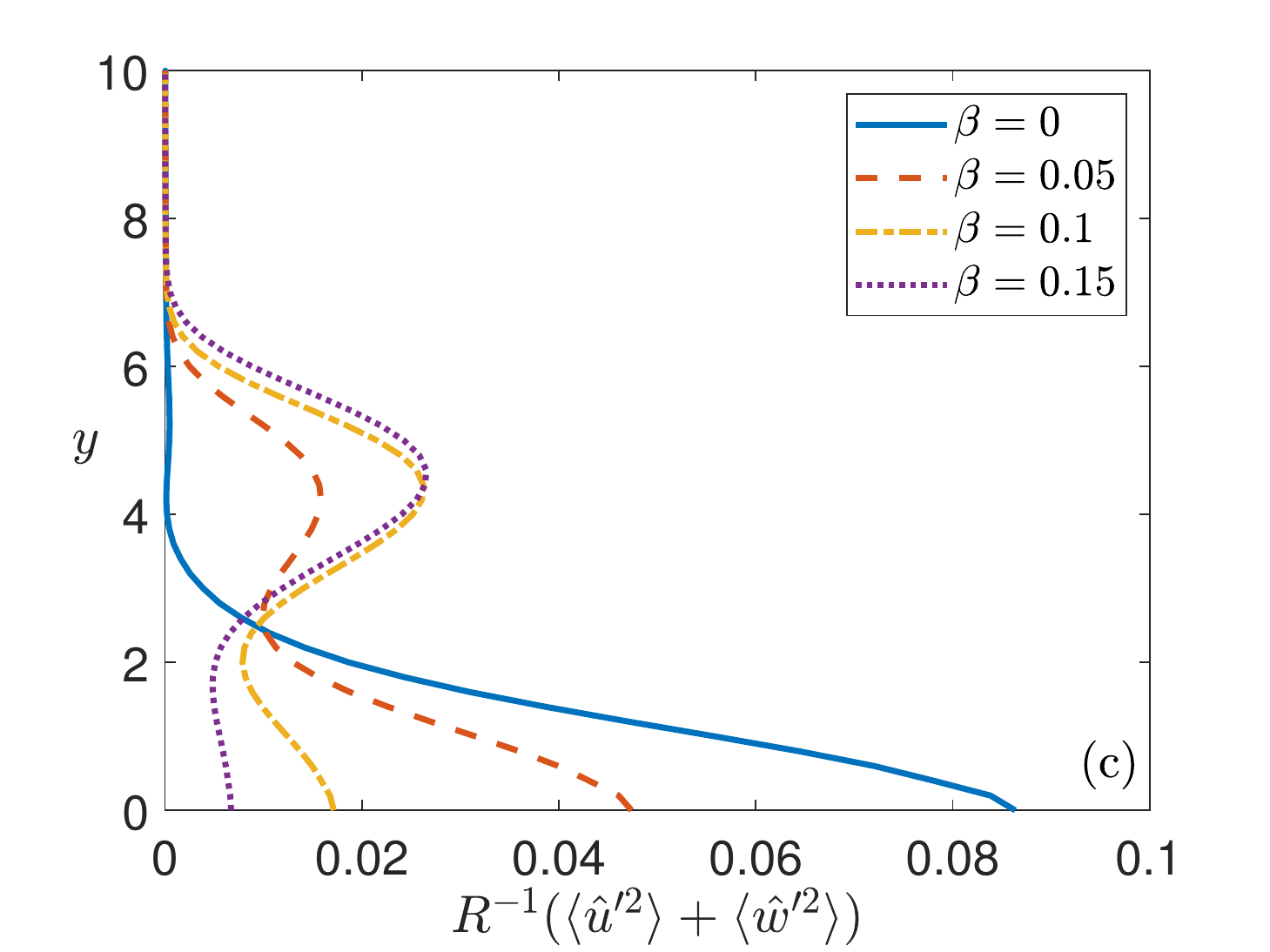}
\caption{\label{fig:EPRS_EDSS_EDV} Plots of the functions associated with the Energy Production due to Reynolds Stresses in $\textrm{(a)}$, the Energy Dissipation due to Surface Stretching in $\textrm{(b)}$, and the Energy Dissipation due to Viscosity in $\textrm{(c)}$. It should be noted that in $\textrm{(c)}$, in the case when $\beta=0$, the plot depends only on $\hat{u}'$ and $R$, since $\hat{w}'\equiv0$. As before, the value of the Reynolds number is fixed at $R=1\times10^{5}$, and the most unstable eigenmode is selected.}
\end{figure}

Before discussing our results associated with the integral energy analysis derived in \ref{subsec:form3} it proves useful to first determine the form of each of the three eigenfunctions. In Fig.~\figref[(a)]{eigs}, Fig.~\figref[(b)]{eigs} and Fig.~\figref[(c)]{eigs} we present plots of $|\hat{u}|$, $|\hat{v}|$ and $|\hat{w}|$ for a range of spanwise wavenumbers. As the value of $\beta$ increases we observe that the streamwise eigenfunctions decay to zero closer to the boundary-layer wall, providing supporting evidence for the observed stabilisation. In addition to this, we also notice that the peak of the wall-normal eigenfunction is also notably reduced. Whilst these eigenfunctions do decay to zero in a region close to $y=y_{\textrm{max}}$, suitable numerical testing has taken place to ensure that these results are indeed independent of our numerical scheme. With regards to the spanwise eigenfunction, no such clear and obvious trends appear to exist. In fact, it would appear that there exists some $\beta$ in the range $0.05<\beta<0.15$ such that the value of $\textrm{max}|\hat{w}|$, is itself maximised. This suggests that the energy production and dissipation terms associated with this system will not increase and decrease linearly as the value of the spanwise wavenumber is increased.

\begin{figure}[t!]
\centering
\includegraphics[width=85mm]{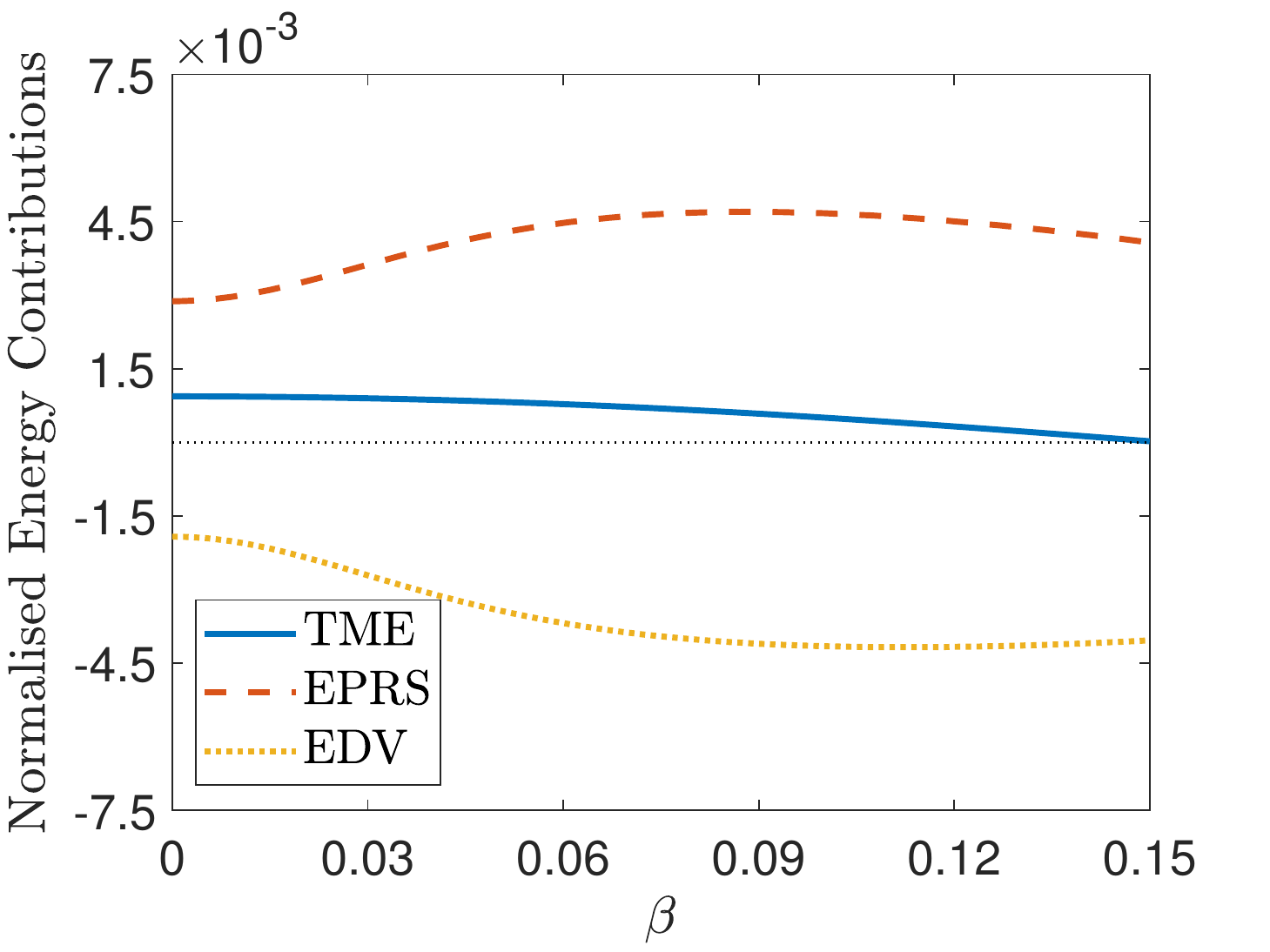}
\caption{\label{fig:energy_plots} A plot of the variation of the Total Mechanical Energy (TME), the Energy Production due to Reynolds Stresses (EPRS) and the Energy Dissipation due to Viscosity (EDV) against the spanwise wavenumber. The point at which the flow transitions from linearly unstable to linearly stable is indicated here by the dotted black line. For all the $\beta$ values presented here the flow is linearly unstable.}
\end{figure}

In Fig.~\ref{fig:EPRS_EDSS_EDV} we then plot the functions associated with the energy production due to Reynolds stresses, and the energy dissipation due to surface stretching and viscosity, for a range of $\beta$ values in order to determine their relative influence on the normalised energy balance. We observe in Fig.~\figref[(a)]{EPRS_EDSS_EDV} that the area encompassed by the energy production curves initially increases as the value of $\beta$ increases. However, in much the same way as before, we notice that the peak of the curve eventually decreases as a function of the spanwise wavenumber. This suggests that there exists a critical value of the spanwise wavenumber, whereby, above this critical value, the contribution of energy production due to Reynolds stresses begins to decrease. It is clear from Fig.~\figref[(b)]{EPRS_EDSS_EDV} that the contribution of energy dissipation from the terms associated with surface stretching are minimal. These terms are much larger in magnitude than those ignored in \ref{subsec:form3}, however, they are still roughly two orders of magnitude smaller than the dissipation term associated with the action of viscosity. It is therefore clear, from this figure, that the additional terms appearing in \eqref{Orr} play a very minimal role in the overall energetics of the system. This suggests that the linear stability of the flow is primarily governed by the form of the base flow and the standard Orr-Sommerfeld equation. It is clear, from Fig.~\figref[(c)]{EPRS_EDSS_EDV}, that the energy dissipation functions associated with the action of viscosity decay to zero farther from the boundary-layer wall as the spanwise wavenumber is increased. This would suggest that the absolute value of the energy dissipation term will likely increase with $\beta$. 

In Fig.~\ref{fig:energy_plots} the normalised energy contributions, as defined in \eqref{energy}, are presented for increasing values of the spanwise wavenumber. Note that here we exclude the contribution from the $\textrm{EDSS}$ terms as these have now been shown to be negligible in this context. As predicted by our prior analysis the energy production initially increases before levelling off and eventually decreasing. The absolute value of the energy dissipation due to viscosity follows a similar trend and, as such, the total mechanical energy of the system decreases with increasing $\beta$. This result is entirely consistent with the conclusions drawn from our neutral stability curve predictions.

\section{\label{sec:asym}Large Reynolds number asymptotic analysis}

In order to investigate the structure of the TS waves in the near-wall viscous layer we now present a large Reynolds number lower branch asymptotic stability analysis. The most amplified TS disturbances appear near to the lower branch of the neutral curve and this, along with the need to validate our numerical solutions, provides the motivation for the lower rather than the upper branch analysis.

We analyse the Orr-Sommerfeld-type equation \eqref{Orr} in the limit as $R\to\infty$, for neutrally stable solutions. Initially, we focus on two-dimensional disturbances ($\beta=0$), so to leading order this becomes Rayleigh's equation, namely

\begin{equation*}
(U-c)(\mathcal{D}^{2}-\alpha^{2})\hat{v}-\hat{v}\mathcal{D}^{2}U=0.
\end{equation*}

\noindent We note that although \eqref{Orr} includes additional terms associated with surface stretching, in the limit as $R\to\infty$, we do recover the standardised version of Rayleigh's equation. This equation holds away from fixed walls and the critical layer (where $U=c$), since, in these regions, viscous effects cannot be ignored. In these regions, for $R\gg 1$, \eqref{Orr} becomes

\begin{equation}
\mathcal{D}^{4}\hat{v}\approx \textrm{i}\alpha R(U-c)\mathcal{D}^{2}\hat{v}.\label{viscous}
\end{equation}

\noindent Close to the wall, $U(y)\approx 1-y+\cdots$, and we find from (\ref{viscous}) that the thickness of the wall layer is $\mathcal{O}((\alpha R(1-c))^{-1/2})$, where we note, from the numerical solutions, that  $0<c<1$. The critical layer is located at $y=y_c$, say, where $U(y_c)=c$, i.e. where $\textrm{e}^{-y_c}=c$, yielding $y_c=-\ln c$. Analysis of (\ref{viscous}) determines the thickness of the critical layer to be $\mathcal{O}((R\omega)^{-1/3})$. On the lower branch of the neutral stability curve, the wall layer and the critical layer merge, yielding $(1-c)\sim(\alpha R)^{-1/3}$. In addition, the numerical solutions suggest that $(1-c)\sim R^{-1/4}$, which leads to the scales $\alpha\sim\omega\sim R^{-1/4}$, on the lower branch of the neutral stability curve. For the case of three-dimensional disturbances $\beta\sim\alpha$, so $\beta\sim R^{-1/4}$, also.

For the ensuing asymptotic analysis it is convenient to non-dimensionalise lengths with respect to a reference length $l^*$, velocities with respect to $\xi^*l^*$, time with respect to $1/{\xi^*}$ and pressure with respect to $\rho^* {\xi^*}^2 {l^*}^2$. Thus, we write $(x^*,y^*,z^*)=l^*(\overline{x},\overline{y},\overline{z})$, $(\tilde{U}^{*},\tilde{V}^{*},\tilde{W}^{*})=\xi^*l^*(\overline{U},\overline{V},\overline{W})$, $t^*=\overline{t}/\xi^*$ and $\tilde{P}^{*}=\rho^* {\xi^*}^2 {l^*}^2 \overline{P}$. This leads to the definition of a Reynolds number $Re=\xi^* {l^*}^2/\nu^*$. We perform a local stability analysis about the streamwise location $\overline{x}_s$ for $Re\gg 1$. The relationship between this Reynolds number and the local Reynolds number used in the numerical analysis is then $R=Re^{1/2}\overline{x}_s$. We will return to this relationship when comparing the asymptotic solutions with the numerical ones. Thus, in terms of $Re$, $(1-c)\sim \alpha\sim\omega\sim Re^{-1/8}$.

The ratio of the length scales is $l^*/L^*=Re^{1/2}$ and similarly, the ratio of the time scales is $Re^{1/2}\overline{x}_s$. This gives the relations $(\overline{x},\overline{y},\overline{z})=Re^{-1/2}(x,y,z)$ and $\overline{t}=Re^{-1/2}t/\overline{x}_s$. Then, in terms of $Re$, the streamwise and spanwise length scales and the timescale, are $\mathcal{O}(Re^{-3/8})$. Also, the thickness of the wall layer is $\mathcal{O}(Re^{-1/2}Re^{-1/8})=\mathcal{O}(Re^{-5/8})$. We introduce scaled coordinates and time to reflect these scales. For convenience, we set $\eps=Re^{-1/8}$, and write

\begin{equation*}
\overline{x}-\overline{x}_s=\eps^3 X, \quad \overline{z}=\eps^3 Z,\quad\textrm{and}\quad \overline{t}=\eps^3\tau.
\end{equation*}

\noindent Thus the non-dimensional disturbed flow is described as such

\begin{equation*}
(\overline{U},\overline{V},\overline{W},\overline{P})
=(\overline{x}U(y),Re^{-1/2}V(y),0,Re^{-1}P(y))
+(\ut,\vt,\wt,\pt),
\end{equation*}

\noindent where the small disturbance quantities, denoted with a tilde, are functions of $\overline{x}$, $\overline{y}$, $\overline{z}$ and $\overline{t}$.

\subsection{Triple-deck structure}

Similarly to the instability analysis of the Blasius flow (uniform flow past a flat plate), we find we have a triple-deck structure, with upper, main and lower deck thicknesses of order $Re^{-3/8}$, $Re^{-1/2}$ and $Re^{-5/8}$, respectively. The analysis turns out to be similar to the corresponding case for a Blasius boundary layer (see Smith \cite{Smith1979}), although with the notable difference that the timescale that the disturbances develop over is equal to the streamwise lengthscale. The main deck covers the extent of the boundary layer, where the disturbances are inviscid and rotational. The upper deck is inviscid and irrotational in nature and required to reduce the disturbances to zero in the far field. The lower deck is required to satisfy the viscous no-slip boundary conditions at the moving surface.

We consider normal-mode solutions and take the perturbations proportional to $E=\exp (\textrm{i}(\theta(X)+\overline{\beta} Z-\overline{\omega}\tau))$. Then the wavenumber $\theta$, is a slowly varying function of $\overline{x}$ in the form

\begin{equation*}
\frac{\textrm{d} \theta}{\textrm{d} X}=\overline{\alpha}=\alpha_0(\overline{x})+\eps\alpha_1(\overline{x})+\cdots.
\end{equation*}

\noindent The spanwise wavenumber $\overline{\beta}$ and the frequency $\overline{\omega}$ are constant and expand as $\overline{\beta}=\beta_0+\eps\beta_1+\cdots$, and $\overline{\omega}=\omega_0+\eps\omega_1+\cdots$. We begin our analysis in the main deck where $\overline{y}=Re^{-1/2}y=\eps^4y$, and the disturbances expand as

\begin{eqnarray*}
\ut&=&(U_{m0}+\eps U_{m1}+\cdots)E,\\
\vt&=&(\eps V_{m0}+\eps^2 V_{m1}+\cdots)E,\\
\wt&=&(\eps W_{m0}+\eps^2 W_{m1}+\cdots)E,\\
\pt&=&(\eps P_{m0}+\eps^2 P_{m1}+\cdots)E,
\end{eqnarray*}

\noindent where $U_{m0},\ldots,V_{m0},\ldots,W_{m0},\ldots$, and $P_{m0}\ldots$, are functions of $y$ and the slow variable $\overline{x}$. We substitute this form for the disturbance quantities into the linear perturbation equations and equate leading order terms. We find that the solutions for the leading-order velocity and pressure disturbances are

\begin{subequations}
\begin{eqnarray}
U_{m0}&=&-B_0(\overline{x})U'(y),\label{eqn:3dUm0}\\
V_{m0}&=&\textrm{i}\frac{B_0(\overline{x})}{\overline{x}_s}(\alpha_0 \overline{x}_s U(y)-\omega_0),\label{eqn:3dVm0}\\
W_{m0}&=&-\frac{\beta_0 P_{m0}(\overline{x})}{\alpha_0 \overline{x}_s U(y)-\omega_0},\label{eqn:3dWm0}\\
P_{m0}&=&P_{m0}(\overline{x}),\label{eqn:3dPm0}
\end{eqnarray}
\end{subequations}

\noindent where $P_{m0}(\overline{x})$ and $B_0(\overline{x})$ are unknown, slowly varying, amplitude functions (representing pressure and negative displacement perturbations, respectively). For these viscous instability modes we choose

\begin{equation*}
\alpha_0 \overline{x}_s =\omega_0.\label{eqn:3dcond1}
\end{equation*}

\noindent This corresponds to the critical layer, where $U=\overline{\omega}/(\overline{\alpha} \overline{x}_s)$, moving to the wall at leading order. An outer layer (the upper deck) is then required to reduce the disturbances to zero as $y\to\infty$.

In the upper deck $\overline{y}=\eps^3 \yt$, with $\yt=\mathcal{O}(1)$.  Of interest here is the matching of the normal velocities between the upper and the main deck. Here

\begin{equation*}
\vt=(\eps \olv_0+\eps^2 \olv_1+\cdots)E,
\end{equation*}

\noindent and similarly for $\ut$, $\wt$ and $\pt$. In the upper deck the basic flow has the behaviour $U\to 0$, and $V\to -1$. The solution for $\olv_0$ is found to be

\begin{equation*}
\olv_0=\textrm{i}\frac{(\alpha_0^2+\beta_0^2)^{1/2}}{\alpha_0 \overline{x}_s}P_{m0}(\overline{x}) \textrm{e}^{-(\alpha_0^2+\beta_0^2)^{1/2} \yt}.
\end{equation*}

\noindent Matching $\olv_0$ as $\yt\to 0$, with $V_{m0}$ as $y\to\infty$, yields the relation

\begin{equation}
B_0(\overline{x})=-\frac{(\alpha_0^2+\beta_0^2)^{1/2}}{\alpha_0^2 \overline{x}_s}P_{m0}(\overline{x}).\label{BP}
\end{equation}

\noindent The desired dispersion relation is obtained by matching the solutions in the main deck with those in the lower deck, which is examined next.

In the lower deck $\overline{y}=\eps^5 \olY$, with $\olY=\mathcal{O}(1)$. Close to the wall $U\approx 1-\eps \olY+\cdots$ and $V\approx -\eps\olY+\cdots$. Then, to match with the main deck solutions, the disturbance quantities expand as

\begin{eqnarray*}
\ut&=&(U_0+\eps U_1+\cdots)E,\\
\vt&=&(\eps^2 V_0+\eps^3 V_1+\cdots)E,\\
\wt&=&(W_0+\eps W_1+\cdots)E,\\
\pt&=&(\eps P_0+\eps^2 P_1+\cdots)E,
\end{eqnarray*}

\noindent where $U_{0},\ldots,V_{0},\ldots,W_{0},\ldots,$ and $P_{0},\ldots$ depend on $\olY$ and $\overline{x}$.

Leading order terms in the $y$-momentum equation yield $P_0=P_{m0}$. The remaining equations show that $\alpha_0{U_0}_{\olY}+\beta_0{W_0}_{\olY}$, satisfies Airy's equation, namely,

\begin{equation}
{ (\alpha_0{U_0}_{\olY}+
\beta_0{W_0}_{\olY})}_{\olY\olY}
-(\textrm{i}(\alpha_1 \overline{x}_s-\omega_1)-\textrm{i}\alpha_0\overline{x}_s \olY)(\alpha_0{U_0}_{\olY}+
\beta_0{W_0}_{\olY})
=0,\label{eqn:3dairy}
\end{equation}

\noindent where here the subscript $\olY$ denotes differentiation with respect to $\olY$. The solution for $U_0$ must satisfy

\begin{equation*}
U_0(\olY=0)=0,\quad
U_0(\olY\to\infty)\to -\frac{(\alpha_0^2+\beta_0^2)^{1/2}}{\alpha_0^2 \overline{x}_s}P_{m0}.
\end{equation*}

\noindent By setting

\begin{equation*}
\zeta=(-\textrm{i}\alpha_0 \overline{x}_s)^{1/3}\left(\olY-\frac{\alpha_1 \overline{x}_s-\omega_1}{\alpha_0 \overline{x}_s}\right),
\end{equation*}

\noindent then the solution of \eqref{eqn:3dairy}, which is bounded as $\zeta\to\infty$, is

\begin{equation*}
\alpha_0{U_0}_{\zeta}+\beta_0{W_0}_{\zeta}=C_0 \textrm{Ai}(\zeta),
\end{equation*}

\noindent where $\textrm{Ai}$ is the (appropriately decaying) Airy function and the subscript $\zeta$ denotes differentiation with respect to $\zeta$. Integrating the above and satisfying the boundary conditions $U_0(\zeta=\zeta_0)=W_0(\zeta=\zeta_0)=0$, yields

\begin{equation*}
\alpha_0U_0+\beta_0W_0=C_0\int_{\zeta_0}^\zeta \textrm{Ai}(\overline\zeta)\, \textrm{d} \overline\zeta,
\end{equation*}

\newpage

\noindent where

\begin{equation*}
\zeta_0=\textrm{e}^{\textrm{i}5\pi/6}(\alpha_0 \overline{x}_s)^{-2/3}(\alpha_1 \overline{x}_s-\omega_1).
\end{equation*}

\noindent Applying the boundary conditions $U_0(\olY=0)=V_0(\olY=0)=0$, in the leading-order $y$- and $z$-momentum equations
yields

\begin{equation*}
{U_0}_{\olY\olY}(\olY=0)-\textrm{i}\alpha_0 P_0={W_0}_{\olY\olY}(\olY=0)-\textrm{i}\beta_0 P_0=0.
\end{equation*}

\noindent Thus, we determine the following relation between $C_0$ and $P_{m0}$

\begin{equation}
(-\textrm{i}\alpha_0 \overline{x}_s)^{2/3}C_0 \textrm{Ai}'(\zeta_0)=\textrm{i} (\alpha_0^2+\beta_0^2)P_{m0}.\label{eqn:3dc1}
\end{equation}

\subsection{Eigenrelation}

Matching $\alpha_0U_0+\beta_0W_0$ between the main and lower decks, using (\ref{BP}), gives a second relation between $C_0$ and $P_{m0}$

\begin{equation}
C_0\int_{\zeta_0}^\infty \textrm{Ai}(\zeta)\, \textrm{d} \zeta=-\frac{(\alpha_0^2+\beta_0^2)^{1/2}}{\alpha_0 \overline{x}_s}P_{m0}.\label{eqn:3dc2}
\end{equation}

\noindent Combining \eqref{eqn:3dc1} and \eqref{eqn:3dc2} yields the desired eigenrelation

\begin{equation}
\textrm{Ai}'(\zeta_0)
=\kappa(-\textrm{i}\alpha_0 \overline{x}_s)^{1/3}(\alpha_0^2+\beta_0^2)^{1/2},\label{eqn:eig3d2}
\end{equation}

\noindent where $\kappa=\int_{\zeta_0}^\infty \textrm{Ai}(\zeta)\, \textrm{d} \zeta$. It is possible to scale $\overline{x}_s$ from the above eigenrelation by writing

\begin{equation*}
(\alpha_0,\beta_0)=\overline{x}_s^{-1/4}(\ola_0,\olb_0),\quad\textrm{and}\quad
\alpha_1\overline{x}_s-\omega_1=\overline{x}_s^{1/2}\olg_1.
\end{equation*}

\noindent The eigenrelation \eqref{eqn:eig3d2} then becomes

\begin{equation}
\textrm{Ai}'(\zeta_0)
=\kappa \textrm{e}^{-\textrm{i}\pi/6}\ola_0^{1/3}(\ola_0^2+\olb_0^2)^{1/2},\label{eqn:eig3d3}
\end{equation}

\noindent where $\zeta_0=\textrm{e}^{\textrm{i}5\pi/6}\ola_0^{-2/3}\olg_1$.

\begin{figure}[t!]
\centering
\includegraphics[width=85mm]{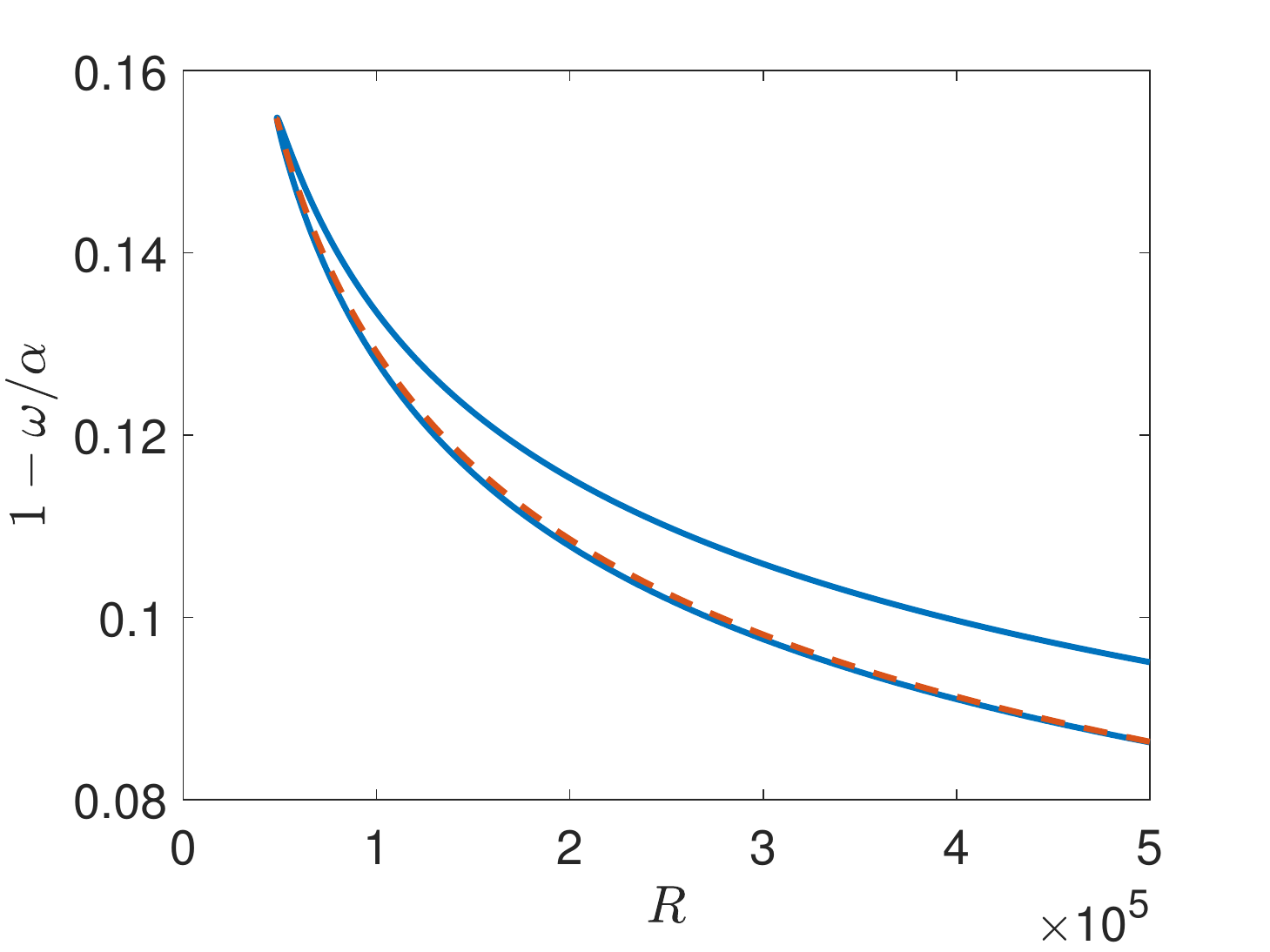}
\caption{Comparison of the asymptotic approximation $1-\omega/\alpha\approx2.296R^{-1/4}$, with our numerical result in the case when $\olb_0=\beta=0$. The solid line represents the curve of neutral stability for two-dimensional disturbances. The dashed line is the lower branch asymptotic solution.}
\label{figa1}
\end{figure}

This eigenrelation can be expressed in terms of the Tietjens function (see, for example, \citet{Reid1965}). Using the notation in \citet{Healey1995} this function is given as

\begin{equation*}
F^{+}(\xi_0)=1-\frac{\textrm{Ai}'(\xi_0)}{\xi_0\int_{\infty}^{\xi_0} \textrm{Ai}(\xi)\,\textrm{d}\xi},
\end{equation*}

\noindent where $\xi_0=\textrm{e}^{-\textrm{i}5\pi/6}z$. For our problem of the stability of the flow due to a linear stretching sheet, the eigenrelation \eqref{eqn:eig3d3} becomes

\begin{equation}
F^+(\zeta_0)-1=-\frac{\ola_0}{\olg_1}(\ola_0^2+\olb_0^2)^{1/2}.\label{eqn:eig3d4}
\end{equation}

\noindent Restricting our attention to neutrally stable solutions ($\ola_0$ real), since $\zeta_0$ is the complex conjugate of $\xi_0$, inspection reveals that $F^{+}(\zeta_0)$ is the complex conjugate of $F^{+}(\xi_0)$. A well-known property of the Tietjens function $F^{+}(\textrm{e}^{-\textrm{i}5\pi/6}z)$ is that it is real for $z=z_0\approx 2.297$, with $F^{+}(\textrm{e}^{-\textrm{i}5\pi/6}z_0)\approx 0.564$. In relation to the instability of the Blasius boundary layer, $z=z_0$ corresponds to the lower branch of the neutral stability curve, while $F^{+}(\xi_0)\to 0$, as $|\xi_0|\to\infty$, and this limit is relevant to the upper branch (\citet{Healey1995}). Thus, approximations to the neutral solutions of \eqref{eqn:eig3d4} for two-dimensional disturbances can be obtained easily, yielding $\ola_0\approx 1.001$, and $\olg_1\approx 2.299$. These values are confirmed by numerical solution of \eqref{eqn:eig3d4}.

\begin{figure}[t!]
\centering
\includegraphics[width=75mm]{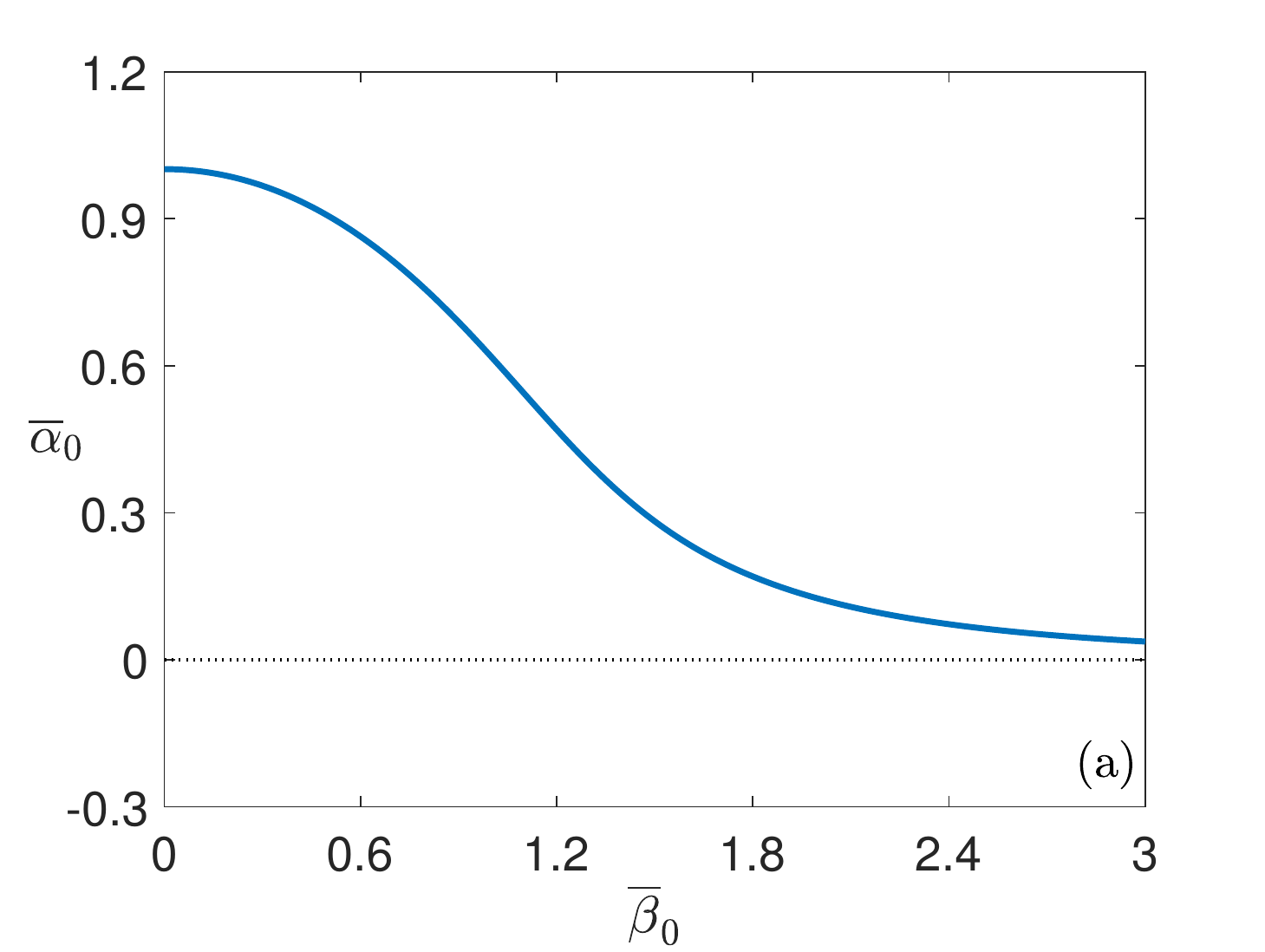}
\includegraphics[width=75mm]{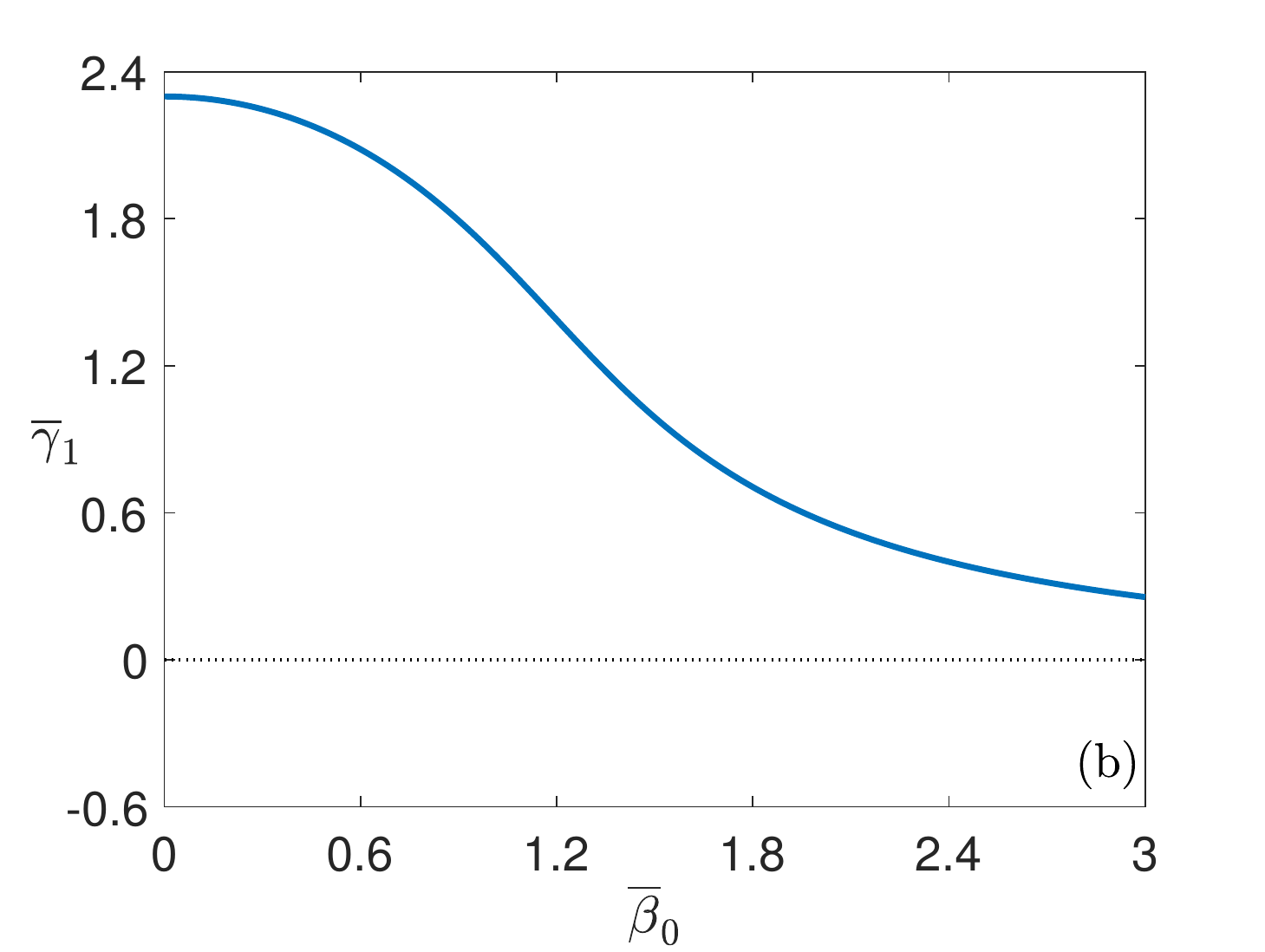}
\caption{Plots of the neutral solutions of \eqref{eqn:eig3d4} for a range of $\olb_0$ values. In $\textrm{(a)}$ and $\textrm{(b)}$ we plot $\ola_0$ and $\olg_1$ against $\olb_0$, respectively.}
\label{figa2}
\end{figure}

In order to compare the asymptotic predictions with the numerical solutions obtained in \ref{sec:num_results} we consider an asymptotic expansion in terms of  $c=\omega/\alpha$. We have that

\begin{equation}
1-\frac{\omega}{\alpha}=1-\frac{\overline{\omega}}{\overline{\alpha}\overline{x}_s}=\eps\frac{\alpha_1 \overline{x}_s-\omega_1}{\alpha_0 \overline{x}_s}+\cdots= R^{-1/4}\frac{\olg_1}{\ola_0}+\cdots.\label{eqn:compR}
\end{equation}

\noindent Thus, for $\olb_0=0$, this gives the approximation $1-\omega/\alpha\approx 2.296R^{-1/4}$. This is compared with the numerical solution from \ref{sec:num_results} in Fig.~\ref{figa1}, showing excellent agreement. Solutions of \eqref{eqn:eig3d4} can also be obtained numerically for $\olb_0\ne 0$. The neutral values of $\ola_0$ and $\olg_1$ obtained for a range of values of $\olb_0$ are shown in Fig.~\ref{figa2}. The first term in the approximation of $R^{1/4}(1-\omega/\alpha)=\olg_1/\ola_0$, can then be inferred for a range of values of $\olb_0$.

\begin{figure}[t!]
\includegraphics[width=75mm]{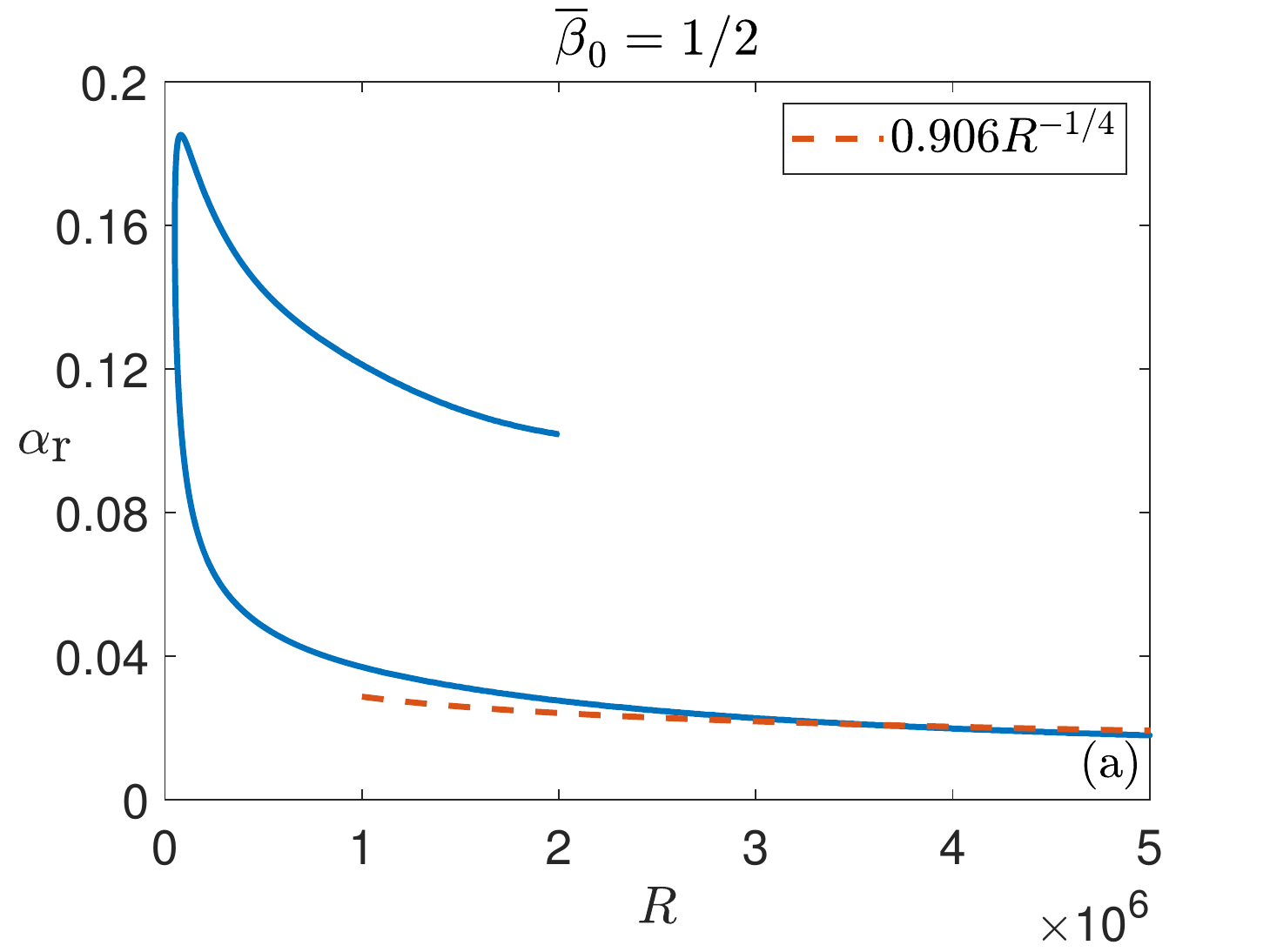}
\includegraphics[width=75mm]{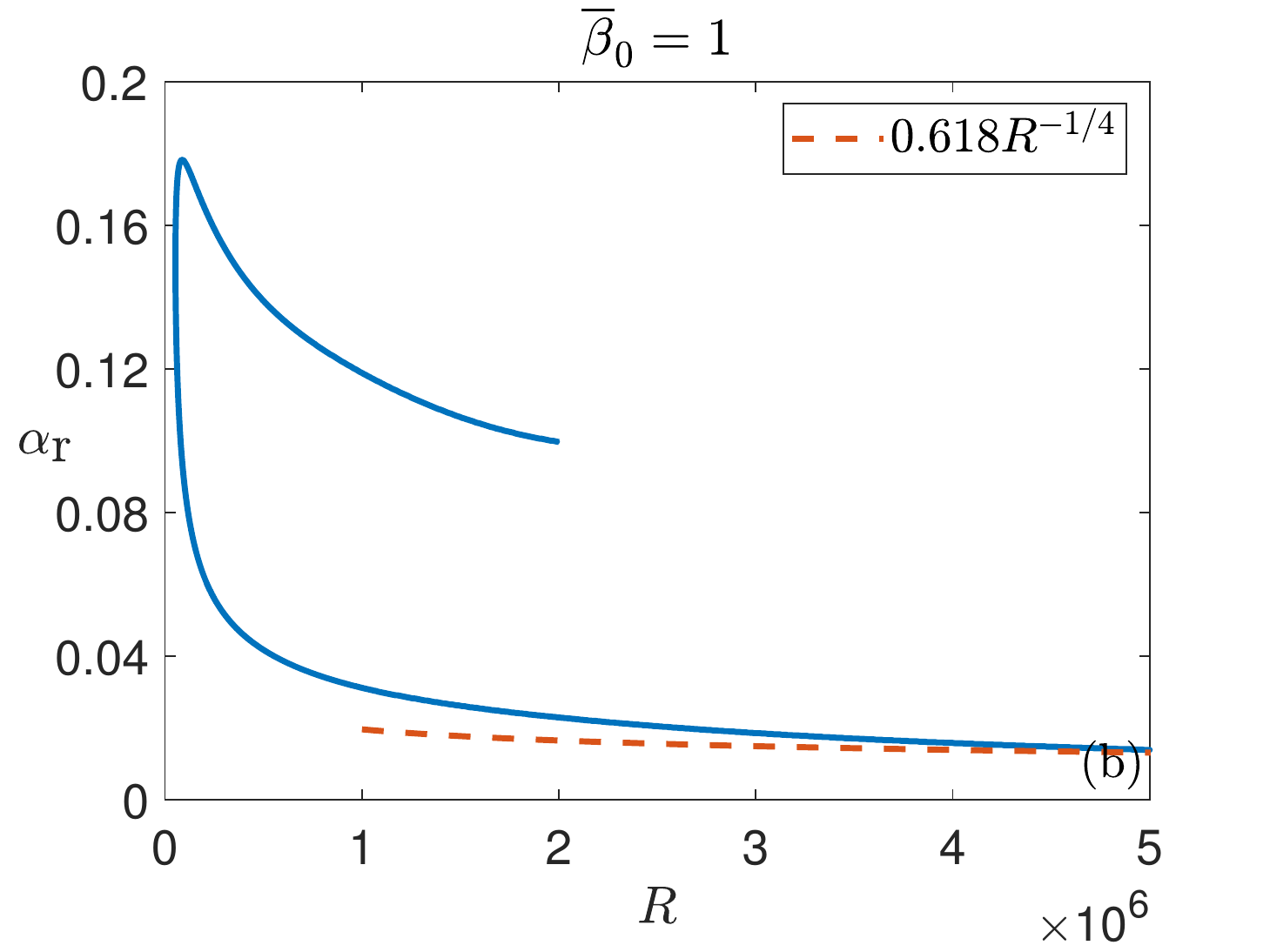}
\caption{\label{fig:NSC_asym_comp} Comparison between the three-dimensional asymptotic solutions and the corresponding curves of neutral stability. The results for the cases when $\olb_0=1/2$, and $\olb_0=1$ are plotted in ($\textrm{a}$) and ($\textrm{b}$), respectively. At each step in the numerical solution procedure the value of the spanwise wavenumber is updated according to the relation $\beta=R^{-1/4}\olb_0$. The upper branch of the neutral stability curve has been truncated at $R=2\times10^{6}$.}
\end{figure}

In order to compare the three-dimensional results with the computational results we use the relation $\beta=R^{-1/4}\olb_0$
at leading order. Then the asymptotic solution for a fixed value of $\olb_0$ can be compared with the computational solution for three-dimensional modes where $\beta$ is given as above. For each value of $R$ the value of $\beta$ is determined and the neutral values of $\alpha$ and $\omega$ are determined (a similar comparison is made for the asymptotic suction boundary layer by \citet{DempseyWalton2017} in their Fig.~3). In Fig.~\ref{fig:NSC_asym_comp} we plot such comparisons for the cases when $\olb_0=1/2$, and $\olb_0=1$. As before excellent agreement can be seen between the one-term asymptotic result and lower branch of the numerically calculated neutral stability curve.

\section{\label{sec:disc}Discussion and Conclusions}

We have assessed the onset of instability of the flow due to a linear stretching sheet. Previous studies, which consider only G\"{o}rtler-type disturbances, have concluded that this flow is linearly stable. However, our analysis has revealed that this flow is susceptible to disturbances in the form of Tollmien-Schlichting waves. 

The flow itself is a rare example of an exact analytical solution of the Navier-Stokes equations. Our analyses, both numerical and analytical, reveal that this flow is linearly unstable to travelling wave disturbances. Although we have been unable to rigorously prove an equivalent to Squire's theorem for this type of flat plate boundary-layer flow we have shown numerically that 2D disturbances are indeed the most unstable. In this case, the critical Reynolds number is of the same order as other flat plate flows exhibiting exponentially decaying base flow solutions.

Our integral energy analysis reveals that the total mechanical energy of the system will always decrease as the spanwise wavenumber increases. This result aligns with the fact that the critical Reynolds number increases (with associated decreasing growth rates) as the value of $\beta$ increases from zero. Given that the streamwise velocity component is $x$-dependent and that the wall-normal velocity component is non-zero, a number of additional terms appear in a governing Orr-Sommerfeld-type equation. However, these terms are all of order $\mathcal{O}(R^{-1})$, and, as evidenced by our integral energy analysis, therefore play a negligible role in the linear stability characteristics of this system. 

The large Reynolds number asymptotic analysis presented in \ref{sec:asym} produces excellent results when compared to our numerical solutions. Of particular note is the exceptional agreement observed between the two sets of solutions in the most unstable case when $\overline{\beta}_{0}=\beta=0$. 

Given that for this type of flow, the local Reynolds number $R$ is directly related to the streamwise location at which the stability analysis is applied, one can infer the dimensional lengthscale associated with the onset of linear instability given that the critical Reynolds number, stretching rate and kinematic viscosity are known

\begin{equation*}
x_{s}^{*}=R\frac{\nu^{*}}{\xi^{*}L^{*}}=R\sqrt{\frac{\nu^{*}}{\xi^{*}}}.
\end{equation*}

Given the critical Reynolds number results presented here, with a fluid of kinematic viscosity $\nu^{*}=1\times10^{-6}\,\textrm{m}^{2}\textrm{s}^{-1}$, and a dimensional stretching rate of $\xi^{*}=20\,\textrm{s}^{-1}$ (as is consistent with the analysis of \citet{Vleggaar1977}, in fact, this estimate for $\xi^{*}$ is likely to be very conservative given the data presented in Fig.~1 of Vleggaar's study), the onset of linear instability for a flow involving the cooling of a continuously extended sheet would be predicted to be of the order of $x_{s}^{*}=\mathcal{O}(10^{1})\,\textrm{m}$. As such, the possibility of this type of instability being realised in an industrial context are perhaps somewhat limited. However, in order to be able to more accurately model industrial processes involving surface stretching a number of additional considerations should be taken in to account. Including the effects of a large temperature gradient on the flow of a polymeric fluid would result in a modification of not only the base flow profiles but also the governing perturbation equations (as evidenced in related studies by \citet{Milleretal2018} and \citet{Craccoetal2020}, for example). This, in turn, would affect the predictions for the critical Reynolds number for the onset of linear instability. Combining this with the fact that the intrinsic properties of the fluid would also be changing, could result in a significantly reduced prediction for $x_{s}^{*}$. Indeed, this will be the goal of the continuation of this study. Using the framework developed here we plan to include the above effects so as to ascertain the relative importance of T-S wave disturbances in a more industrially relevant flow settings. Having said that, this study has clearly been successful in that we have been able to clearly demonstrate that flows of this nature are indeed susceptible to disturbances in the form of Tollmien-Schlichting waves, confirming the earlier speculation of \citet{BhattacharyyaGupta1985}.

\section*{Acknowledgements}
PTG and MK would like to acknowledge the generous support of the School of Mathematics and Statistics at The University of Sydney where part of this study was completed.

\section*{Data Availability Statement}

The data that support the findings of this study are available from the corresponding author upon reasonable request.

\end{document}